\documentclass[fleqn,usenatbib]{mnras}

\usepackage{newtxtext,newtxmath}

\usepackage[T1]{fontenc}

\DeclareRobustCommand{\VAN}[3]{#2}
\let\VANthebibliography\thebibliography
\def\thebibliography{\DeclareRobustCommand{\VAN}[3]{##3}\VANthebibliography}

\usepackage{graphicx}
\usepackage{bm}

\usepackage{tikz,xcolor,hyperref}

\definecolor{lime}{HTML}{A6CE39}
\DeclareRobustCommand{\orcidicon}{%
	\begin{tikzpicture}
	\draw[lime, fill=lime] (0,0) 
	circle [radius=0.16] 
	node[white] {{\fontfamily{qag}\selectfont \tiny ID}};
	\draw[white, fill=white] (-0.0625,0.095) 
	circle [radius=0.005];
	\end{tikzpicture}
	\hspace{-2mm}
}

\foreach \x in {A, ..., Z}{%
	\expandafter\xdef\csname orcid\x\endcsname{\noexpand\href{https://orcid.org/\csname orcidauthor\x\endcsname}{\noexpand\orcidicon}}
}

\def \bb {\bm{b}}

\def \bj {\bm{j}}
\def \br {\bm{r}}

\title[Dissipation measures in weakly collisional plasmas]{Dissipation in weakly collisional plasmas}

\author[O. Pezzi et al.]{
O. Pezzi\orcidA{}$^{1,2,3}$\thanks{\textrm{Email: oreste.pezzi@gssi.it}}, 
$^{}$H. Liang\orcidB{}$^{4}$, 
J.~L. Juno\orcidC{}$^{5}$,  
P.~A. Cassak\orcidD{}$^{6}$, 
C.~L. V\'asconez\orcidE{}$^{7}$, 
L. Sorriso-Valvo\orcidF{}$^{8,3}$, \newauthor 
D. Perrone\orcidG{}$^{9}$,
S. Servidio\orcidH{}$^{10}$, 
V. Roytershteyn\orcidI{}$^{11}$, 
J.~M. TenBarge\orcidJ{}$^{12}$ \& 
W.~H. Matthaeus\orcidK{}$^{13}$\\ \\
$^{1}$Gran Sasso Science Institute, Viale F. Crispi 7, I-67100 L'Aquila, Italy\\
$^{2}$INFN/Laboratori Nazionali del Gran Sasso, I-67100 Assergi (AQ), Italy\\
$^{3}$Istituto per la Scienza e Tecnologia dei Plasmi, Consiglio Nazionale delle Ricerche, Via Amendola 122/D, I-70126 Bari, Italy\\
$^{4}$Center for Space Plasma and Aeronomic Research, University of Alabama in Huntsville, Huntsville, AL 35899, USA\\
$^{5}$Department of Physics and Astronomy, University of Iowa, Iowa City, IA 54224, USA\\
$^{6}$Department of Physics and Astronomy and Center for KINETIC Plasma Physics, West Virginia University, WV 26506, USA\\
$^{7}$Departamento de F\'isica, Escuela Polit\'ecnica Nacional, Ladr\'on de Guevara E11-253, 170525 Quito, Ecuador\\
$^{8}$Swedish Institute of Space Physics, \r{A}ngstr\"om Laboratory, L\"agerhyddsv\"agen 1, SE-751 21 Uppsala, Sweden\\
$^{9}$ASI -- Italian Space Agency, via del Politecnico snc, I-00133 Rome, Italy\\
$^{10}$Dipartimento di Fisica, Universit\`a della Calabria, I-87036 Rende (CS), Italy\\
$^{11}$Space Science Institute, Boulder, CO 80301, USA\\
$^{12}$Department of Astrophysical Sciences, Princeton University, Princeton, NJ 08544, USA\\
$^{13}$Bartol Research Institute and Department of Physics and Astronomy, University of Delaware, Newark, DE 19716, USA
}

\date{Accepted 2021 May 21; Revised 2021 May 21; Received 2021 January 4}
\pubyear{2021}

\begin{document}
\label{firstpage}
\pagerange{\pageref{firstpage}--\pageref{lastpage}}
\maketitle

\begin{abstract}
The physical foundations of the dissipation of energy and the associated heating in weakly collisional plasmas are poorly understood. Here, we compare and contrast several measures that have been used to characterize energy dissipation and kinetic-scale conversion in plasmas by means of a suite of kinetic numerical simulations describing both magnetic reconnection and decaying plasma turbulence. We adopt three different numerical codes that can also include interparticle collisions: the fully kinetic particle-in-cell {\sc vpic}, the fully kinetic continuum \texttt{Gkeyll}, and the Eulerian Hybrid Vlasov--Maxwell (HVM) code.
We differentiate between (i) four {\it energy}-based parameters, whose definition is related to energy transfer in a fluid description of a plasma, and (ii) four {\it distribution function}-based parameters, requiring knowledge of the particle velocity distribution function. There is an overall agreement between the dissipation measures obtained in the PIC and continuum reconnection simulations, with slight differences due to the presence/absence of secondary islands in the two simulations. There are also many qualitative similarities between the signatures in the reconnection simulations and the self-consistent current sheets that form in turbulence, although the latter exhibits significant variations compared to the reconnection results. All the parameters confirm that dissipation occurs close to regions of intense magnetic stresses, thus exhibiting local correlation. The distribution function-based measures show a broader width compared to energy-based proxies, suggesting that energy transfer is co-localized at coherent structures, but can affect the particle distribution function in wider regions. The effect of interparticle collisions on these parameters is finally discussed. 
\end{abstract}



\section{Introduction}
\label{sect:intro}

Understanding energy dissipation and heating in weakly collisional plasmas is a key challenge in the study of space and astrophysical plasmas, such as the solar corona, solar wind, the outer magnetospheres of planets, and compact astrophysical objects. At variance with neutral fluids and collisional plasmas (e.g. magnetofluids), interparticle collisions are typically weak in these systems and are often neglected. One example is during magnetic reconnection, where magnetic fields with a reversing component effectively break and cross-connect at length-scales at and below the gyroradius of the particles (kinetic scales), allowing the conversion of magnetic energy to kinetic, thermal, and non-thermal energy \citep{Yamada:etal:2010RvMP}. Another example is that weakly collisional plasmas are often observed to be in a strongly turbulent state \citep{federrath2010comparing,bruno2016turbulence,narita2018spacetime,beresnyak2019MHD,fraternale2019magnetic}. Hence, they are characterized by the cross-scale transfer of fluctuation energy from large injection scales to smaller, kinetic scales, where energy dissipation is expected to occur \citep{schekochihin2009astrophysical, howes2015dynamical, matthaeus2015intermittency, vaivads2016turbulence,verscharen2019multiscale}. These features have, at least, two profound implications.

First, the dynamics of weakly collisional plasmas involve the whole phase-space, both in configuration and velocity space, as opposed to collisional systems for which the proximity to local thermal equilibrium restricts velocity distribution functions to be near-Maxwellians due to efficient collisional thermalization. Indeed, VDFs in weakly collisional plasmas frequently display non-equilibrium features, such as temperature anisotropy and agyrotropy, rings, beams of accelerated particles, etc. \citep{marsch2006kinetic, servidio2012local, servidio2015kinetic, lapenta2017origin, perri2020deviation}. In turbulence, this interesting dynamics has been envisioned as a phase-space cascade \citep{tatsuno2009nonlinear, plunk2011energy, KanekarEA15, ParkerEA16, SchekochihinEA16, servidio2017magnetospheric}, where the plasma exhibits a power-law scaling --typical of turbulence-- in both physical and velocity space. In this perspective, by expanding particle VDFs in velocity space using Hermite polynomials, a power-law Hermite spectrum was recently observed, reflecting the presence of fine velocity-space structures, in both {\it in situ} observations \citep{servidio2017magnetospheric} and numerical simulations \citep{cerri2018dual, pezzi2018velocityspace}. This clearly illustrates the necessity of using kinetic models to describe such plasmas. Moreover, non-Maxwellian structures in particle VDFs are observed to be significant in the vicinity of intense current sheets or other nearby coherent structures such as vortices, where MHD-like dissipation is thought to occur \citep{osman2011evidence, servidio2012local,matthaeus2015intermittency, parashar2016propinquity}.

Secondly, concepts about energy dissipation and conversion in neutral fluids and magnetofluids do not necessarily carry over to weakly collisional systems. The rate and parametric dependence of dissipation and kinetic-scale energy conversion for such plasmas is extremely challenging and is not thoroughly understood. One challenge is that various physical mechanisms and processes can cause dissipation and kinetic scale energy conversion in various physical circumstances. In weakly collisional plasmas in which kinetic effects play a crucial role in the dynamics, such as during turbulence and reconnection, it may be difficult to assess quantitatively the relative importance of the various dissipative mechanisms, and likewise the net effect of a given dynamical process may be unclear due to reversibility. To address these issues different mechanisms and concepts of dissipation have been proposed and, hence, several dissipation surrogates have been adopted \citep[see e.g.][for recent reviews]{vaivads2016turbulence, matthaeus2020pathways}. However, a unified and general picture of when and where different views work better is still lacking. 

Another important question is the following: do the sites identified potential sites of dissipation correspond to regions where interparticle collisions, although weak, dissipate energy in an irreversible way? Addressing this question has encouraged the examination of a different concept of dissipation associated with the growth of entropy due to collisions \citep{tenbarge2013collisionless, navarro2016structure, pezzi2019protonproton,liang2019decomposition}. Although collisions typically act on large characteristic times \citep{spitzer1956physics,vafin2019coulomb}, their effects are enhanced where particle VDFs exhibit strong distortions, since intense velocity-space gradients are dissipated very quickly by collisions \citep{landau1936transport, rosenbluth1957fokkerplanck, balescu1960irreversible, schekochihin2009astrophysical, pezzi2016collisional, pezzi2017solarwind}. Non-Maxwellian VDFs make intraspecies collision operators non-zero, thus activating this dissipation channel. 

In the following, we classify eight different dissipation proxies into two general classes that we call ``energy-based'' and ``VDF-based''. The energy-based definition describes dissipation as a transfer of energy within a fluid-like description. Often such transfer occurs from an ordered component (e.g. magnetic or bulk flow fluctuations) into a random (e.g., internal) component. On the other hand, VDF-based surrogates directly quantify the presence of non-equilibrium features in the particle VDF. The two classes of dissipation proxies are correlated, since the distortion of the particle VDF is often the consequence of a transfer of energy and vice versa. The dissipation proxies here adopted do not explicitly distinguish signatures associated with particular phenomena, e.g. Landau damping, cyclotron damping, or stochastic processes \citep{ChandranEA10, numata2015ion, li2015dissipation, chen2019evidence}. Future studies will analyze the connection between these dissipation measures --useful to detect potential sites of inhomogeneous dissipation in a turbulent environment-- and the underlying plasma processes, e.g. highlighted through the field-particle correlation \citep{klein2016measuring,klein2017diagnosing,chen2019evidence,klein2020diagnosing}.

In this work, we investigate numerically the functionality of several dissipation proxies belonging to the two classes introduced above. We focus on two different types of numerical simulations: magnetic reconnection in a single current sheet and the development of a turbulent cascade at kinetic scales. We exploit three different numerical Boltzmann--Maxwell algorithms, of both Lagrangian and Eulerian type, that can include inter-particle collisions. In particular, we adopt the fully kinetic particle-in-cell {\sc vpic} code \citep{bowers2008ultrahigh}, and two Eulerian Vlasov--Maxwell codes. These latter codes are the fully kinetic continuum Vlasov--Maxwell solver implemented in the \texttt{Gkeyll} simulation framework \citep{juno2018discontinuous} and the Hybrid Vlasov--Maxwell (HVM) code with kinetic protons and fluid electrons \citep{valentini2007hybrid}. We find that the dissipation measures well characterize significant features of both magnetic reconnection and turbulence, such as the reconnection diffusion region and the intermittent current sheets surrounding turbulent vortices. The dissipation surrogates evaluated from the PIC and \texttt{Gkeyll} reconnection simulations agree to a wide extent. Slight differences between the two runs result from the presence of secondary islands in the \texttt{Gkeyll} simulation that are not present in the PIC simulation. A qualitative correspondence between the signatures in the reconnection simulations and the self-consistent current sheets generated in turbulence is also found, despite larger variations observed in the turbulent case with respect to the magnetic reconnection one. 
The parameters show a {\it regional correlation}: their local peaks take place in similar spatial regions, but they are not necessarily point-to-point correlated \citep{yang2018scale, matthaeus2020pathways}. When including interparticle collisions, peaks of the dissipation proxies are in general weaker than in the associated collisionless system, suggesting that the slow yet incessant effect of collisions locally reduce the transfer of energy and the presence of non-Maxwellian features. By considering the effect of both intraspecies and interspecies collisions, we confirm that the former mainly dissipate non-Maxwellian features in the particle VDF, although they may have an indirect effect also on energy transfer through the pressure tensor isotropization \citep{DelSartoEA16}. On the other hand, the latter also significantly affect energy-based parameters. Finally, by adopting a suite of different algorithms and numerical codes, the current work aims at providing a further contribution to the ``turbulence dissipation challenge'' \citep{parashar2015turbulent}, on which several recent efforts have been dedicated
\citep[e.g.][]{pezzi2017colliding, perrone2018fluid, gonzalez2019turbulent}. 

The paper is structured as follows. In Section \ref{sect:dissmeas}, we define and discuss the dissipation measures investigated in the current work. In Section \ref{sect:nummodel}, numerical models and algorithms adopted for the current analysis are described. Sections \ref{sect:MR} and \ref{sect:turb} report numerical results obtained in the simulations of reconnection and the onset of kinetic turbulence, respectively. In Section \ref{sect:1dcuts}, we show one-dimensional (1D) profiles of the dissipation proxies close to the reconnecting current sheet and a typical current sheet observed in the turbulence simulation. Finally, conclusions and discussions are presented in Section \ref{sect:concl}.

\section{Dissipation measures in weakly collisional plasmas}
\label{sect:dissmeas}
In this section, we introduce the framework and summarize the dissipation proxies adopted in this work. We consider a weakly collisional plasma, composed of protons ($p$) and electrons ($e$). The Boltzmann--Maxwell equations, describing non-relativistic plasmas, in cgs units are:
\begin{align}
&\frac{\partial f_\alpha}{\partial t} + \bm{v} \cdot \frac{\partial f_\alpha}{\partial \bm{r} } +  \frac{q_\alpha}{m_\alpha} \left ( \bm{E} + \frac{\bm{v}}{c} \times \bm{B} \right) \cdot \frac{\partial f_\alpha}{\partial \bm{v}} = \left. \frac{\partial f_\alpha}{\partial t} \right|_{\rm coll}
\label{eq:vlas}\\
&\bm{\nabla} \cdot \bm{E} = 4 \pi \rho_c \label{eq:divE}\\
&\bm{\nabla} \cdot \bm{B} = 0 \label{eq:divB}\\
& \bm{\nabla} \times \bm{E} =  - \frac{1}{c} \frac{\partial \bm{B}}{\partial t} \label{eq:far} \\
& \bm{\nabla} \times \bm{B} =   \frac{1}{c} \frac{\partial \bm{E}}{\partial t} +  \frac{4\pi}{c} \bm{j} \, , 
\label{eq:amp}
\end{align}
where $f_\alpha ({\bm r}, {\bm v}, t) $ is the $\alpha$-species VDF ($\alpha=p,e$); ${\bm E}({\bm r}, t)$ and ${\bm B}({\bm r}, t)$ are the electric and magnetic fields; and $q_\alpha$, $m_\alpha$, and $c$ are the $\alpha$-species charge, mass and the light speed. The charge and electric current densities are, respectively, $\rho_c=\sum_\alpha q_\alpha n_\alpha$ and ${\bm j}=\sum_\alpha q_\alpha n_\alpha {\bm u}_\alpha$, where $n_\alpha = \int d^3v f_{\alpha}$ is the $\alpha$-species number density and $n_\alpha {\bm u}_\alpha = \int d^3v \bm{v} f_{\alpha}$ is the $\alpha$-species number flux density. In the following subsections, we omit the collisional operator $\left. \partial f_\alpha/ \partial t \right|_{\rm coll}$ for simplicity.

\subsection{Energy-based dissipation measures} 
\label{sec-energymeasures}
The energy-based dissipation proxies can be introduced from the energy equations:
\begin{align}
&\frac{\partial \mathcal{E}^f_\alpha}{\partial t} + \nabla \cdot \left({\bm u}_\alpha \mathcal{E}^f_\alpha + {\bm u}_\alpha\cdot {\bm P}_\alpha \right) = \left({\bm P}_\alpha \cdot \nabla \right)\cdot{\bm u}_\alpha + n_\alpha q_\alpha{\bm u}_\alpha \cdot {\bm E}  \label{eq:enflow}\\
&\frac{\partial \mathcal{E}^{th}_\alpha}{\partial t} + \nabla \cdot \left({\bm u}_\alpha \mathcal{E}^{th}_\alpha + {\bm h}_\alpha \right) = -\left({\bm P}_\alpha \cdot \nabla \right)\cdot {\bm u}_\alpha \label{eq:enth}       \\
& \frac{\partial \mathcal{E}^{m}}{\partial t} + \frac{c}{4\pi} \nabla \cdot \left({\bm E} \times {\bm B}  \right) =  - {\bm j} \cdot {\bm E} \label{eq:enmagn} 
\end{align}

where $\mathcal{E}^f_\alpha= \rho_\alpha {\bm u}_\alpha^2/2$ is the bulk flow energy density of the $\alpha$-species, $\mathcal{E}^{th}_\alpha= m_\alpha \int d^3v f_\alpha \left({\bm v}-{\bm u}_\alpha\right)^2/2$ is the thermal (internal) energy density of the $\alpha$-species and $\mathcal{E}^{m}= ({\bm E}^2+{\bm B}^2)/8\pi$ is the electromagnetic energy density. In the above equations, 
\begin{align}
 & {\bm P}_\alpha = \frac{m_\alpha}{2} \int d^3v ({\bm v}-{\bm u}_\alpha) ({\bm v}-{\bm u}_\alpha) f_\alpha \label{eq:Presstens} \\  
 & {\bm h}_\alpha = \frac{1}{2} \int d^3v ({\bm v}-{\bm u}_\alpha)^2 ({\bm v}-{\bm u}_\alpha) f_\alpha \label{eq:heatflux}
\end{align}
are the pressure tensor and vector heat flux, respectively.

The left-hand sides of Eqs. (\ref{eq:enflow})--(\ref{eq:enmagn}) contain terms involving divergences of fluxes. These terms can be locally important and correspond physically to energy transport \citep{pezzi2019energy}. Assuming there is no flux across the domain boundaries (e.g., with periodic boundary conditions), they have no net effect on the global energy partition of the system. Energy transfer between bulk flow and magnetic energy is described by the ${\bm j} \cdot {\bm E}$ term, while conversion of energy between bulk flow and thermal occurs through the pressure--strain interaction $\left({\bm P}_\alpha \cdot \nabla \right)\cdot {\bm u}_\alpha$. Including intraspecies collisional effects (e.g. proton--proton and electron--electron) would not introduce an extra term in Eqs. (\ref{eq:enflow})--(\ref{eq:enmagn}). However these collisions indirectly affect these equations by thermalizing the particle VDF, thus reducing the nongyrotropic terms in the pressure tensor and, in turn, having an impact on the pressure-strain interaction term \citep{DelSartoEA16}. On the other hand, interspecies collisions (electron--proton) insert an explicit interspecies energy transfer term in the energy equations.

The first dissipation surrogate here considered is the Zenitani measure \citep{zenitani2011new}, widely adopted to describe dissipation in magnetic reconnection \citep{zenitani2011new, phan2018electron} and plasma turbulence \citep{wan2015intermittent}. It evaluates the rate of work per unit volume done by the electric field on particles, ${\bm j} \cdot {\bm E}$, in the reference frame co-moving with the considered species, in contrast with ${\bm j} \cdot {\bm E}$ in the simulation frame, which gives the total rate of energy conversion between the electromagnetic fields and the plasma. It directly measures the non-ideal energy conversion  \citep{zenitani2011new} and is related to the production of entropy density in the MHD framework \citep{birn2005energy}. It reads as:
\begin{equation}
    D_{\alpha} = {\bm j}' \cdot {\bm E}' = {\bm j} \cdot \left({\bm E} + \frac{{\bm u}_{\alpha}}{c} \times {\bm B} \right) - \rho_c \left({\bm u}_\alpha \cdot {\bm E}\right) \, ,
    \label{eq:zenitani}
\end{equation}
where ${\bm j}'$ and ${\bm E}'$ are the current density and the electric field in the reference frame co-moving with the species $\alpha$, respectively. As shown in \citet{zenitani2011new}, $n_eD_e=n_pD_p$ and, in a singly ionized quasi-neutral system such as the ones considered in the present work, $D_e \simeq D_p$. $D_\alpha$ contains both reversible and irreversible contributions since the electric field ${\bm E}$ has contributions from both reversible (e.g. wave--particle interactions etc.) and irreversible (e.g. collisional resistive) processes. We note that the probability distribution function (PDF) of $D_\alpha$ and other dissipation measures are almost symmetric between negative and positive values (see Fig. 4 of \citet{wan2015intermittent}). Net dissipation, which is ultimately an integral over space and time, arises from slight asymmetry (skewness) in the tails of the PDF.

More recently, the pressure--strain interaction $\left({\bm P}_{\alpha} \cdot \nabla \right)\cdot {\bm u}_{\alpha}$ has been analysed to understand dissipative mechanisms in weakly collisional plasmas \citep{yang2017energyPOP,yang2017energyPRE,yang2018scale,chasapis2018insitu,sitnov2018kinetic,pezzi2019energy,matthaeus2020pathways}. This term is commonly decomposed as
\begin{equation}\label{eq:pid}
 -\left({\bm P}_{\alpha} \cdot \nabla \right)\cdot {\bm u}_{\alpha} = - P_{\alpha} \theta_{\alpha} - {\bm \Pi}_{\alpha} : \boldsymbol{\mathcal{D}}_{\alpha} \, ,
\end{equation}
where $P_{{\alpha},ij} = P_{\alpha} \delta_{ij} + \Pi_{{\alpha},ij}$; $P_{\alpha}=P_{{\alpha},ii}/3$; $\theta_{\alpha}=\nabla \cdot {\bm u}_{\alpha}$; $\mathcal{D}_{{\alpha},ij}=\left(\partial_j u_{{\alpha},i} + \partial_i u_{{\alpha},j} \right)/2 - \theta_{\alpha} \delta_{ij}/3$; and $\delta_{ij}$ and $\partial_i$ denote the Kronecker delta and a partial derivative with respect to the $i$--th spatial coordinate, respectively. The first term on the right-hand side of Eq. (\ref{eq:pid}), called P-$\theta_\alpha$, is associated with plasma expansion and compression. The last term, called Pi-D$_\alpha$, is associated with the trace-less (anisotropic and off-diagonal) parts of the pressure tensor and the symmetric part of the velocity strain and describes the rate of work per unit volume done by flow shear. The spatial integral of the Pi-D$_\alpha$ term measures the thermal energy gain \citep{pezzi2019energy}. Here, we use the convention that Pi-D$_\alpha$ and P-$\theta_\alpha$ include the minus signs in Eq. (\ref{eq:pid}), so that positive values tend to locally increase the thermal energy. The Pi-D$_\alpha$ term has been adopted to provide insights on the mechanisms that transfer energy towards smaller scales, where it is dissipated \citep{yang2018scale}. The motivation for this term being associated with dissipation arises from the MHD framework, in which traceless pressure--tensor terms are related to viscous dissipation \citep{braginskii1965transport}. Hence, it is interesting to explore whether this connection remains valid in a weakly collisional system.

Another dissipation proxy is associated with the cross-scale conversion of energy in its non-linear transfer during the turbulent cascade~\citep{Frisch95}. A proxy of the scale-dependent local energy transfer rate (LET) in a weakly collisional plasma at proton inertial scales may be estimated through the combined velocity, magnetic field and current fluctuations as~\citep{sorriso2018local,sorriso2018statistical,sorriso2019turbulence}
\begin{equation}
\begin{split}
 \epsilon_{p,\ell}  =  
  \left(|\Delta {\bm u}_{p}|^2 + |\Delta \bb|^2\right) \frac{\Delta u_{p,\ell}} {\ell} - 2 (\Delta {\bm u}_{p} \cdot \Delta \bb) \frac{\Delta b_\ell} {\ell} \\
  - \frac{d_p}{2} |\Delta \bb|^2 \frac{\Delta j_\ell} {\ell} +  d_p (\Delta \bb \cdot \Delta \bj) \frac{\Delta b_\ell} {\ell}  \, 
\end{split}  
\label{let}
\end{equation}
where $d_p$ is the proton skin depth. The magnetic field is in velocity units ${\bm b} \equiv {\bm B}/\sqrt{4\pi\rho}$, where $\rho$ is the proton mass density. $\Delta {\bm g} \equiv {\bm g}(\br + {\bm \ell}) - {\bm g} (\br)$ is the increment of a field ${\bm g}$ between two spatial points separated by a distance ${\bm \ell}$ and is used to describe structures or fluctuations of that size. The subscript $\ell$ indicates longitudinal increment, i.e. the projection along the increment vector ${\bm \ell}$. In this work, LET values are associated with the starting point of the increment vector and, for all simulations, the LET parameter $\epsilon_{p,\ell}$ is computed with $\ell = d_p$ and by averaging the increments taken in the positive horizontal and vertical directions. Since the LET parameter $\epsilon_{p,\ell}$ is introduced within a fluid framework, we do not compute it with $\ell$ below proton  scales, where electron kinetic effects become important. The definition of LET is motivated by the isotropic form of the Politano--Pouquet scaling law for the mixed third-order fluctuations in turbulent incompressible MHD plasmas~\citep{politano1998dynamical,sorriso2002analysis,sorriso2007observation}, in this case also including the Hall--MHD terms~\citep{galtier2008vonkarman,ferrand2019exact,bandyopadhyay2020insitu,vasconez2020local}. Under the assumption of stationarity, homogeneity and large Reynolds' number, the associated scaling coefficient $\langle \epsilon_{p} \rangle$  ($\langle...\rangle $ indicating the ensemble average) is the constant mean energy transfer (or, in the stationary steady state, the dissipation) rate of the turbulent cascade~\citep{Kolmogorov41a}. In 3D fluid, fully developed turbulence, the sign of the averaged third-order moment has to be negative, corresponding to an ensemble-averaged global net transfer towards the small scales (the non-linear turbulent energy cascade). Other terms locally contributing to the energy transfer vanish and are disregarded here. These represent the different fluid contributions to the turbulent transfer of energy: kinetic and magnetic turbulent energy transported by velocity fluctuations, coupled magnetic and velocity Alfv\'enic fluctuations transported by magnetic structures, and two associated Hall terms. 
In weakly collisional plasmas, the LET may be unable to describe the contribution of compressibility, as well as the role of possible non-thermal features. The latter might enter in the energy budget via the pressure--tensor contributions. However, it can provide information on the local transfer of ordered energy towards small scales, where it is made available for conversion through various possible mechanisms, including dissipation. In particular, according to the definition used in this work, negative LET terms can be thought of as locally contributing to the non-linear energy transfer towards small scales, and positive LET terms describe a contribution to energy transfer to large scales. In non-turbulent systems as well as in systems displaying a large-scale structure (e.g. a single reconnecting current sheet), the interpretation of the LET sign is more difficult, since the assumption of homogeneity is not satisfied.  
However, for this analysis, peaks of LET indicate the local concentration of cross-scale energy transfer that can be made available for dissipation, regardless of the sign.

\subsection{VDF-based dissipation measures}
The first parameter belonging to the class of VDF-based dissipation measures is the pressure agyrotropy \citep{scudder2009illuminating}. Agyrotropy is a measure of differences in the plasma temperatures in the two perpendicular directions to a given axis \citep{scudder2009illuminating,aunai2013electron,swisdak2016quantifying}. Standard axis orientations for computing agyrotropy are along the local magnetic field or the mean magnetic field. Other coordinate systems, such as the minimum variance frames of the particle VDF, have been also adopted \citep{servidio2015kinetic,pezzi2017colliding}. We consider here the agyrotropy parameter $\sqrt{Q_\alpha}$ proposed by \citet{swisdak2016quantifying}, evaluated relative to the local magnetic field. Writing the pressure tensor ${\bm P}_{\alpha}$ in the coordinate system where the local magnetic field is parallel to the $z$-axis as
\begin{equation}
{\bm P}_{\alpha}=
 \begin{pmatrix}
P_{\alpha,\perp} & P_{\alpha,12} & P_{\alpha,13}\\
P_{\alpha,12} & P_{\alpha,\perp} & P_{\alpha,23}\\
P_{\alpha,13} & P_{\alpha,23} & P_{\alpha,\parallel}
\end{pmatrix},
\end{equation}
the agyrotropy parameter is defined as
\begin{equation}
    Q_{\alpha} = \frac{P_{\alpha,12}^2 + P_{\alpha,13}^2 + P_{\alpha,23}^2}{P_{\alpha,\perp}^2 + 2 P_{\alpha,\perp}P_{\alpha,\parallel}}.
    \label{eq:Q}
\end{equation}
$Q_{\alpha}$ can be computed in an arbitrary coordinate system, as explained in appendix A of \citet{swisdak2016quantifying}.

The variety of non-Maxwellian structures observed during magnetic reconnection or in a turbulent plasma is much richer than just pressure agyrotropies. Another proxy, usually named $\epsilon$ but here called $\xi$ since $\epsilon$ is already used to indicate the LET proxy, was proposed by \citet{greco2012inhomogeneous}. Here we propose a slightly different, non-dimensional definition (the original definition had dimensions of $v^{-3/2}$):
\begin{equation}  
\xi_{\alpha}({\bm r},t) = \frac{v_{th,\alpha}^{3/2}({\bm r},t)}{n_{\alpha}({\bm r},t)}\sqrt{\int d^3v \left[f_{\alpha}({\bm r},{\bm v}, t)-g_{\alpha}({\bm r},{\bm v},t)\right]^2  } \, .
 \label{eq:eps}
\end{equation}
Here, $g_{\alpha}$ is the equivalent Maxwellian distribution function associated with $f_{\alpha}$, i.e. constructed using the local values of the density $n_\alpha$, bulk speed ${\bm u}_\alpha$ and total temperature $T_\alpha$ of the $\alpha$-species; while $v_{th,\alpha}=\sqrt{k_B T_\alpha/m_\alpha}$ is the (local) thermal speed. 

Similar measures to identify non-Maxwellian VDFs were constructed using kinetic entropy. The kinetic entropy density $s_\alpha$ is
\begin{equation}
 \label{eq:entrdens}
 s_{\alpha}({\bm r}, t) = - k_B \int d^3v f_{\alpha}({\bm r},{\bm v},t) \log{f_{\alpha}({\bm r},{\bm v},t)} \; .
\end{equation}
Note the total entropy $S_\alpha = \int s_\alpha d^3 r$ of a collisional system is non-decreasing from the Boltzmann H-theorem, but the entropy density may locally increase or decrease \citep{pezzi2019protonproton}. It is possible to define a velocity--space entropy density $s^{\rm vel}_{\alpha}$ retaining only the spatially local contribution to entropy from permutations of particles in velocity space \citep{liang2019decomposition} as
\begin{equation}\label{eq:svel}
 s^{\rm vel}_{\alpha} ({\bm r},t) = s_\alpha({\bm r},t) + k_B n_{\alpha}({\bm r},t) \log\left(\frac{n_{\alpha}({\bm r},t)}{\Delta v^3}\right) \;
\end{equation}
where $\Delta^3 v$ is the volume of the cell in velocity space. Using these definitions, two dimensionless non-Maxwellianity parameters have been introduced:
\begin{equation}\label{eq:MKP}
 \bar{M}_{{\rm KP},\alpha} ({\bm r},t) = \frac{s_{\rm M,\alpha}({\bm r},t) - s_{\alpha}({\bm r},t)}{(3/2)k_B n_{\alpha}({\bm r},t)} = \frac{s^{\rm vel}_{{\rm M},\alpha}({\bm r},t) - s^{\rm vel}_{\alpha}({\bm r},t)}{(3/2)k_B n_{\alpha}({\bm r},t)} \, ,
 \end{equation}
proposed by \citet{kaufmann2009boltzmann}, and
\begin{equation}\label{eq:M}
 \bar{M}_\alpha ({\bm r},t) = \frac{\bar{M}_{{\rm KP},\alpha}({\bm r},t)}{1 + \log \left( \frac{2\pi k_B T_\alpha}{m_\alpha(\Delta^3 v)^{2/3}}\right)} \, ,
\end{equation}
proposed by \citet{liang2020kinetic}. Here, $s_{{\rm M},\alpha}({\bm r},t)$ is the entropy density evaluated using the equivalent Maxwellian distribution function $g_\alpha({\bm r},{\bm v},t)$ associated with the VDF $f_\alpha({\bm r},{\bm v},t)$, given by
\begin{equation}
    s_{{\rm M},\alpha} = \frac{3}{2} k_B n_\alpha \left[ 1 + \log\frac{2\pi k_B T_{\alpha}}{m_\alpha n_{\alpha}^{2/3}} \right] \, .
\end{equation}
and $s^{\rm vel}_{{\rm M},\alpha}({\bm r},t)$ is computed according to Eq. (\ref{eq:svel}). 

In a system with a fixed number of particles and total energy, the Maxwellian distribution has the maximum entropy. Hence, $\bar{M}_{\rm KP}$ and $\bar{{M}}$ (along with $\xi$) are positive definite. These parameters measure all higher order VDF disturbances from the local Maxwellian beyond its second-order moment. This retains information about dissipation, since high-order variations from the Maxwellian coincide with the presence of fine velocity--space structures that, in turn, are dissipated by collisional effects \citep{pezzi2016collisional}. 

\section{Numerical models and simulations setup}
\label{sect:nummodel}

The parameters introduced in the previous section are computed using the results of kinetic numerical simulations performed with different codes.  All simulations in this work are 2.5D in space (quantities depend on two dimensions, but vectors have three components) and 3D in velocity space. For all codes, quantities are presented using a normalization based on an arbitrary magnetic field strength $B_0$ and density $n_0$. Spatial and temporal scales are normalized to the proton inertial length $d_p=c/\omega_{pp}$ and the proton cyclotron time $\Omega_{cp}^{-1}$, respectively, where $\omega_{pp}=\sqrt{4\pi n_0e^2/m_p}$ is the proton plasma frequency based on $n_0$ and $\Omega_{cp}=eB_0/m_pc$ is the proton cyclotron frequency based on $B_0$. Thus, velocities are normalized to the Alfv\'{e}n velocity $c_A=d_p \Omega_{cp}$; electric fields are normalized to $c_A B_0/c$; pressures and temperatures are normalized to $B_0^2 / 4\pi$ and $m_p c_A^2/k_B$, respectively; and entropy is normalized to Boltzmann's constant $k_B$ [see \citet{liang2019decomposition} for a detailed discussion of the units of the continuous Boltzmann entropy]. Derived units of the dissipation measures are therefore as follows: $D_\alpha$, Pi-D$_\alpha$, and P-$\theta_\alpha$ are $\Omega_{cp} B_0^2 / 4\pi$, $\epsilon_p$ is $c_A^2 \Omega_{cp}$, while $Q_\alpha, \xi_\alpha, \bar{M}_{KP,\alpha}$, and $\bar{M}_\alpha$ are dimensionless.

\subsection{Numerical algorithms}

For the present analysis, we adopt the particle-in-cell {\sc vpic} code  and two different Eulerian Vlasov--Maxwell codes: the fully kinetic \texttt{Gkeyll} code and the hybrid-kinetic HVM code.

{\sc vpic} utilizes a 3D, relativistic, fully kinetic explicit algorithm \citep{bowers2008ultrahigh}. {\sc vpic} has been widely adopted for both collisionless and weakly collisional plasma simulations, including simulations of magnetic reconnection and plasma turbulence  \citep[e.g.][]{daughton2009collisional,Daughton:etal:2011NatPh, karimabadi2013coherent, Roytershteyn2013collisional3D, wan2015intermittent,Roytershteyn2015holes}. The code includes several models of binary collisions, including the particle-pairing Coulomb collision algorithm of \citet{takizuka1977binary} capable of accurately reproducing the Landau collisional integral over a wide range of parameters. The latter model is used in this study. 

\texttt{Gkeyll} is a highly extensible code framework that contains solvers for a number of systems of equations of relevance to plasma physics, including multimoment multifluid \citep{wang2015fluid}, continuum gyrokinetics \citep{shi2019gk, mandell2020gk}, and continuum Vlasov--Maxwell \citep{juno2018discontinuous,hakimjuno2020supercomputing}. \texttt{Gkeyll}'s Vlasov-Maxwell solver utilizes the discontinuous Galerkin finite element method for phase-space discretization and a strong-stability preserving Runge--Kutta method for the integration in time. The conservative, discontinuous Galerkin implementation of the non-linear Dougherty operator \citep{dougherty1964model} is adopted to include intraspecies collisions \citep{hakim2020collisions} \citep[see][for further details]{juno2020thesis}. 

HVM integrates the Vlasov--Maxwell system within the hybrid framework, assuming quasi-neutrality and neglecting the displacement current density \citep{mangeney2002numerical,valentini2007hybrid}. The proton Vlasov equation is discretized on a phase-space grid and integrated numerically, while electrons are assumed to be a massless isothermal fluid. A generalized Ohm's law for evaluating the electric field in Faraday's law is coupled to the Vlasov equation. Proton--proton collisions have been recently included through the non-linear Dougherty operator \citep{pezzi2015collisional, pezzi2019fourier,pezzi2019protonproton}. 

\subsection{Simulations setup}\label{sec:simulationSetup}

We discuss the two classes of numerical simulations in this work. The first class employs both collisionless and weakly collisional {\sc vpic} and \texttt{Gkeyll} simulations of a single current sheet that undergoes symmetric antiparallel magnetic reconnection. In the collisionless case, we find that {\sc vpic} and a separate PIC code P3D \citep{zeiler2002threedimensional}, adopted in \citet{liang2020kinetic}, provide consistent and qualitatively similar results. Although the {\sc vpic} and \texttt{Gkeyll} simulations are very similar in their choice of parameters, there are small differences we make note of in the subsequent discussion. We first describe the {\sc vpic} simulations.

The {\sc vpic} reconnection simulations use a domain size of $L_x \times L_z=25 \times 25$, with periodic boundary conditions in $x$ and perfectly conducting boundaries on $z$. A single-current-sheet initial condition is used, with magnetic field given by $B_x(z)= \tanh [(z-L_z/2)/w_0] $, where $w_0=0.5$ is the initial half-thickness of the current sheet. The initial VDFs are drifting Maxwellians with temperatures $T_e=1/12$ and $T_p=5/12$ for electrons and protons, respectively; both temperatures are initially uniform over the whole domain. The density is set to balance plasma pressure in the fluid sense, with $n(z)={\rm sech}^2 [(z-L_z/2)/w_0] +n_b$, where $n_b=0.2$ is the background (lobe) density. Therefore, the total upstream plasma $\beta$ for this simulation is $n_b k_B (T_e + T_p) / (B_0^2 /8 \pi) = 0.2$. The proton-to-electron mass ratio is $m_p/m_e=25$ and the speed of light $c=15$. These choices enforce that the plasma is non-relativistic (the thermal and Alfv\'en speeds are much less than the light speed), which is appropriate for the non-relativistic treatment of kinetic entropy. We employ a time-step of $\Delta t \approx 5.8 \times 10^{-4}$. The smallest electron Debye length for this simulation (based on the maximum density of $1 + n_b$) is $\lambda_{De}=0.018$. The spatial grid scale is $\Delta x=\Delta z=0.0125\approx 0.6944 \ \lambda_{De}$ ($N_x=N_z=2000$). The reference number of particles per cell per particle species is $10^{4}$ for a density equal to one. We simulate three different electron--ion collision frequencies: $\nu=0$, $\nu=0.01 \ \Omega_{ce}=0.25 \ \Omega_{cp}$ and $\nu=0.05 \ \Omega_{ce}=1.25 \ \Omega_{cp}$.
All types of collisions (electron--electron, electron--ion, and ion--ion) are taken into account. For each type of collision, the variance of the scattering angle in the Takizuka--Abe algorithm is chosen to yield correct ratio of the respective collision frequencies \citep[see][~for more details]{takizuka1977binary}.

In addition to the parameters for the PIC simulation, the kinetic entropy diagnostic requires other parameters, discussed in detail in appendix B of \citet{liang2019decomposition}. For the PIC simulation, we use a velocity space grid scale of $\Delta v \approx 0.6 \ v_{th,e}$ for electrons and $\Delta v \approx \ 0.5 v_{th,i}$ for ions. We use a velocity range for binning the particles from $-c$ to $c$ in each dimension for electrons and from $-0.4 c$ to $0.4 c$ for ions. 

The \texttt{Gkeyll} reconnection simulation also uses a single current sheet initial condition, in a domain of size, $L_x \times L_z = 8\pi \times 4\pi$, which compared to the PIC simulation is a similar size in $x$ but about half the size in $z$. The boundary conditions are periodic in $x$, a reflecting wall boundary condition in $z$. The initial magnetic fields and plasma parameters are the same as the {\sc vpic} simulations: $w_0 = 0.5$, $m_p/m_e = 25, T_p = 5/12, T_e = 1/12,$ and $n_b = 0.2$. Likewise, there is only one Maxwellian component in the current sheet, but $c=50$. Since the continuum Vlasov method in \texttt{Gkeyll} avoids aliasing errors associated with under-resolving the Debye length in PIC methods while still conserving energy, we choose a coarser configuration space grid resolution to save on the computational cost of a continuum method while still resolving the reconnection dynamics. Our grid resolution is $\Delta x = \Delta z \approx 0.4 \approx 40 \ \lambda_{De}$ ($N_x=64$, $N_z=32$) with piecewise quadratic Serendipity polynomials within a grid cell \citep{arnold2011serendipity}. The velocity space range is from $-6 \ v_{th,\alpha}$ to $6 \ v_{th,\alpha}$ with a velocity space grid of $\Delta v_\alpha = v_{th,\alpha}$ along with piecewise quadratic Serendipity polynomials in velocity space. Zero-flux boundary conditions are employed in velocity space to ensure energy conservation. We choose a constant collisionality of $\nu_{ee} = 0.01$ for electron--electron collisions and $\nu_{pp} = 0.002$ for proton--proton collisions for the Dougherty collision operator. Reconnection is initiated using a magnetic perturbation with a spectrum of random wave modes in the first 20 modes of the system with r.m.s.~amplitude $\delta B/B_0 = 2\times 10^{-3}$. These random perturbations break the symmetry of the continuum kinetic initial condition and allow for the study of the standard $m=1$ tearing mode that arises from noise in PIC simulations.

The HVM turbulence simulations have $512$ grid-points in each direction and a size $L_x=L_y=L=2\pi\times 20$. Periodic boundary conditions are imposed for the spatial domain. Velocity space is discretized with $71$ grid-points in the range $v_{j}=\left[-5 v_{th,p},5v_{th,p}\right]\;(j=x,y,z)$, with the boundary condition $f(v_j>5 v_{th,p})=0$. The initial equilibrium is characterized by spatial homogeneity, Maxwellian proton VDFs, and a background uniform out-of-plane magnetic field ${\bm B_0} = {\bm e}_z$ with $\beta_p=2$. This equilibrium is perturbed at $t=0$ by imposing magnetic $\bm{\delta B}$ and bulk speed $\bm{\delta u}=\pm \bm{\delta b}$ fluctuations ($\bm{\delta b}$ in Alfv\'en speed units). Energy is injected at large scales, i.e. $k\in[2,6]k_0$ ($k_0=2\pi/L$), with a flat energy spectrum and random phases. The r.m.s.~amplitude of the fluctuations is $\delta B/B_0 = 1/2$. No density perturbations nor parallel perturbations are introduced at $t=0$. Electron inertia effects are neglected in Ohm's law, while electron temperature is set equal to the initial ion temperature. A small resistivity ($\eta\simeq10^{-3}$) is introduced to suppress numerical instabilities and does not play a significant role in the plasma dynamics.
The adopted numerical resolution captures two decades of perpendicular wavenumbers: one above and one below the proton skin depth $d_p$. We consider two simulations, characterized by a different proton--proton collisional frequency $\nu$, namely collisionless ($\nu=0$) and  weakly collisional ($\nu=10^{-3}$) \citep[see][for further details]{pezzi2019protonproton}.

\section{Numerical results: Dissipation measures in magnetic reconnection}
\label{sect:MR}

\subsection{Proton dissipation proxies for collisionless reconnection}

\begin{figure*}
\centering
\includegraphics[width=0.9\textwidth]{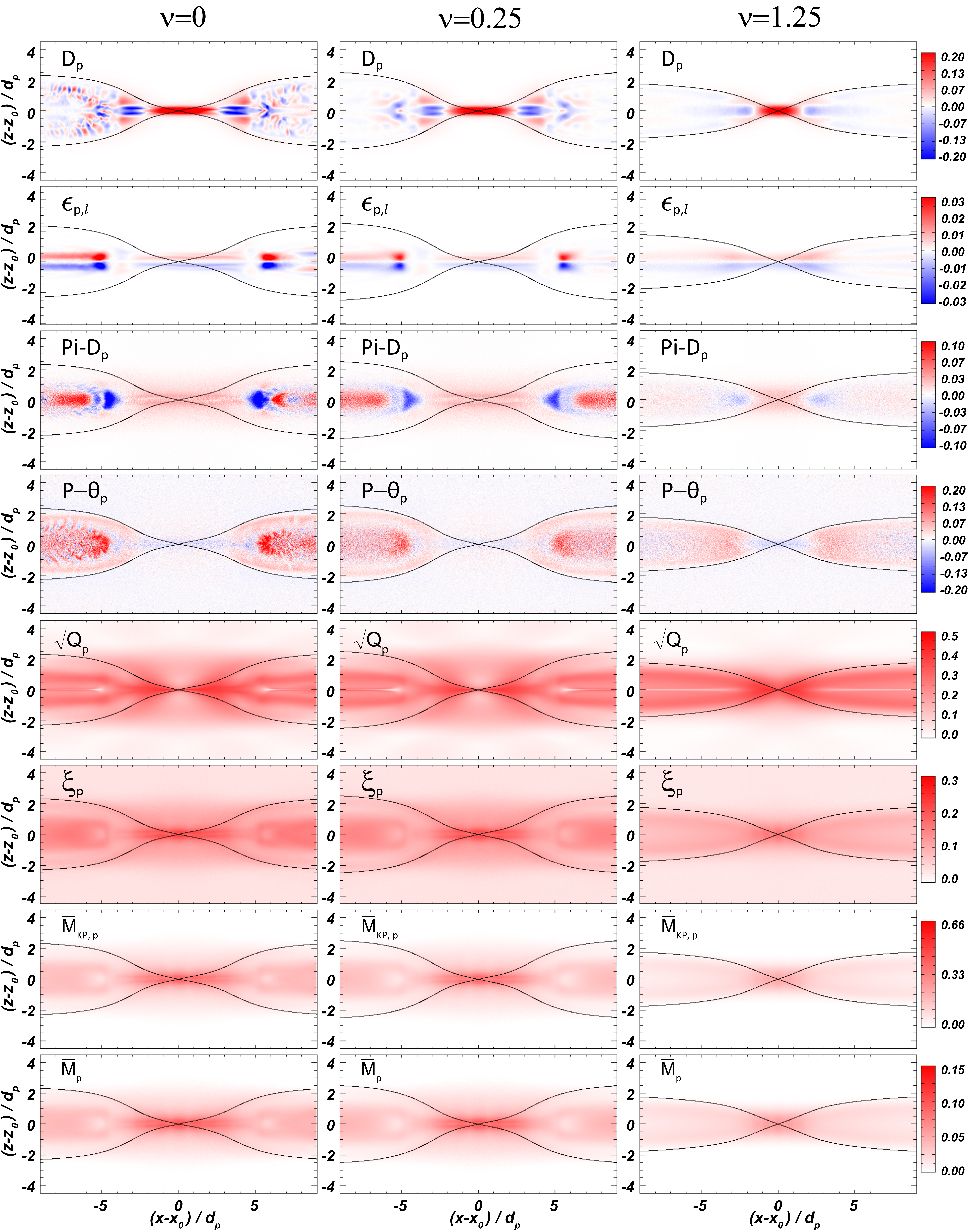}
\caption{Comparison of various dissipation proxies for protons, from the {\sc vpic} simulations with $\nu=0$ (left column), $\nu=0.25$ (centre) and $\nu=1.25$ (right). The proxies are computed at $t=22$ for the collisionless run and $t=26$ for the weakly collisional runs. From top to bottom: the Zenitani parameter $D_p$; the LET parameter $\epsilon_{p,\ell}$ with $\ell\simeq d_p$; the pressure--strain interaction Pi-D$_p$; the pressure dilatation P-$\theta_p$; the agyrotropy $\sqrt{Q}_p$; the non-Maxwellian indicators $\xi_p$; $\bar{M}_{{\rm KP},p}$; and $\bar{M}_p$. Solid lines indicate separatrices.}
  	\label{fig:vpic_i}
\end{figure*}

Figure~\ref{fig:vpic_i} displays the set of implemented proxies for protons in the {\sc vpic} simulations. The left column collects the results for the collisionless simulation, i.e. with $\nu=0$. We initially focus on these results, and discuss the effects of collisions in Section~\ref{sec-collisions}. The data are taken at time $t\simeq 22 \ \Omega_{cp}^{-1}$, after the peak of the reconnection rate. To compute energy-based parameters involving spatial derivatives, Gaussian smoothing is used to filter the noise  \citep[e.g.][]{birdsall2004plasma}.

The Zenitani parameter $D_p$ shows signatures consistent with previous studies \citep{zenitani2011new, swisdak2016quantifying}, being peaked with a positive value in the diffusion region. In the exhausts, there is oscillatory behavior especially on small scales in the primary island. This is likely due to time-domain structures \citep{Mozer15} such as electron holes, which form Debye-scale bi-directional electric fields. Such structures are at small scales and produce local energy conversion between the particles and fields.

The LET parameter $\epsilon_{p,\ell}$ shows a large-scale pattern peaked inside the magnetic island that is only weakly modulated in the horizontal direction, as well as a weaker signal approximately coincident with the electron diffusion region (EDR). As pointed out in Section~\ref{sec-energymeasures}, extracting information about LET is challenging in this reconnection simulation, where the background field is inhomogeneous and the fields are not in a steady state of fully developed turbulence. The signs of LET are opposite on either side of the magnetic reversal because LET is calculated with an increment with a component in the positive $z$ direction.
The contribution of various terms to the local $\epsilon_{p,\ell}$ (not shown) reveals that the Hall terms (in particular the current-helicity term, i.e., the last term in Eq. \ref{let}) dominate the non-linear energy transfer. This is related to the presence of Hall-scale electric currents and to the magnetic configuration of the reconnection region, which thus are the main drivers of the non-linear interactions. 
Furthermore, the prevalent positive sign observed for the MHD cross-helicity term (not shown) suggests that non-linear interactions are inhibited by the strong presence of coupled velocity-magnetic field (Alfv\'enic) fluctuations, which reduce the effective transfer of energy and possibly the onset of turbulence.

The Pi-D$_p$ plot shows that the pressure--strain term is positive yet small near the X point on a length-scale in the inflow direction beyond the EDR. Protons undergo meandering orbits in this region, producing non-gyrotropic VDFs, while the reconnection inflow and outflow are associated with bulk velocity shear: this produces a non-zero Pi-D$_p$. Inside the islands, a bipolar (positive/negative) signal is found. The strong negative region suggests energy is locally being converted from thermal energy to bulk kinetic energy, perhaps in a region where counterstreaming beams including reflected ions at the dipolarization front where the denser current sheet population is being pushed downstream by the reconnected magnetic field are converted into bulk flow.  In contrast, P-$\theta_p$, which has a peak value about a factor of $2$ larger than Pi-D$_p$, tends to be quite structured and positive in most of the island, is negative in the EDR, and is small in the ion diffusion region (IDR). These results make sense physically: the plasma in the island is undergoing compression due to the bulk flows, so that P-$\theta_p$ is positive. In the diffusion region, when upstream magnetic flux tubes enter the region of weaker magnetic field, they expand, leading to negative $\theta_p = \nabla \cdot {\bm u}_p$.
 
Moving to VDF-based parameters, the proton agyrotropy $\sqrt{Q_p}$ parameter indicates a proton gyro-scale region of non-gyrotropy surrounding the diffusion region and the whole island at this stage of the evolution, owing to the complicated distribution functions that appear where protons undergo meandering orbits. Local $\sqrt{Q_p}$ maxima occur in the inner shell of the magnetic island, where both the LET and Pi-D$_p$ are locally peaked. Turning to the non-Maxwellianity parameters, the $\xi_p$ parameter similarly shows structure at proton scales in both the diffusion region and the islands.  The structure of $\bar{M}_{\rm KP,p}$ and $\bar{{M}}_p$ are qualitatively quite similar to $\xi_p$, as expected. In each case, the protons are most strongly non-Maxwellian in the EDR, with non-zero values also in the IDR and in the island. The main difference between these last measures, namely the quadratic non-Maxwellian parameter $\xi_p$ and the entropy-based non-Maxwellianity parameters, is that the latter appear more localized than the former, suggesting that the entropy-based proxies require relatively more strongly non-Maxwellian structures to attain appreciable values relative to $\xi_p$, owing to the natural log VDF dependence in their definition compared to the quadratic dependence of $\xi_p$.

Summarizing the collisionless simulation results for protons, all eight dissipation measures in question show structure in and around the current sheet and magnetic island. For the parameters in this simulation, the quantities that are most strongly peaked in the EDR are $D_p, \xi_p, \bar{M}_{KP,p}$, and $\bar{M}_p$. The strongest measure for the IDR is $\sqrt{Q}_p$, with Pi-D$_p, \xi_p, \bar{M}_{KP,p}$, and $\bar{M}_p$ also displaying structure. In the island, $\epsilon_p,$ Pi-D$_p$ and P-$\theta_p$ are significant, with $D_p$ revealing significant electron-scale variations.

\begin{figure*}
\centering
\includegraphics[width=0.9\textwidth]{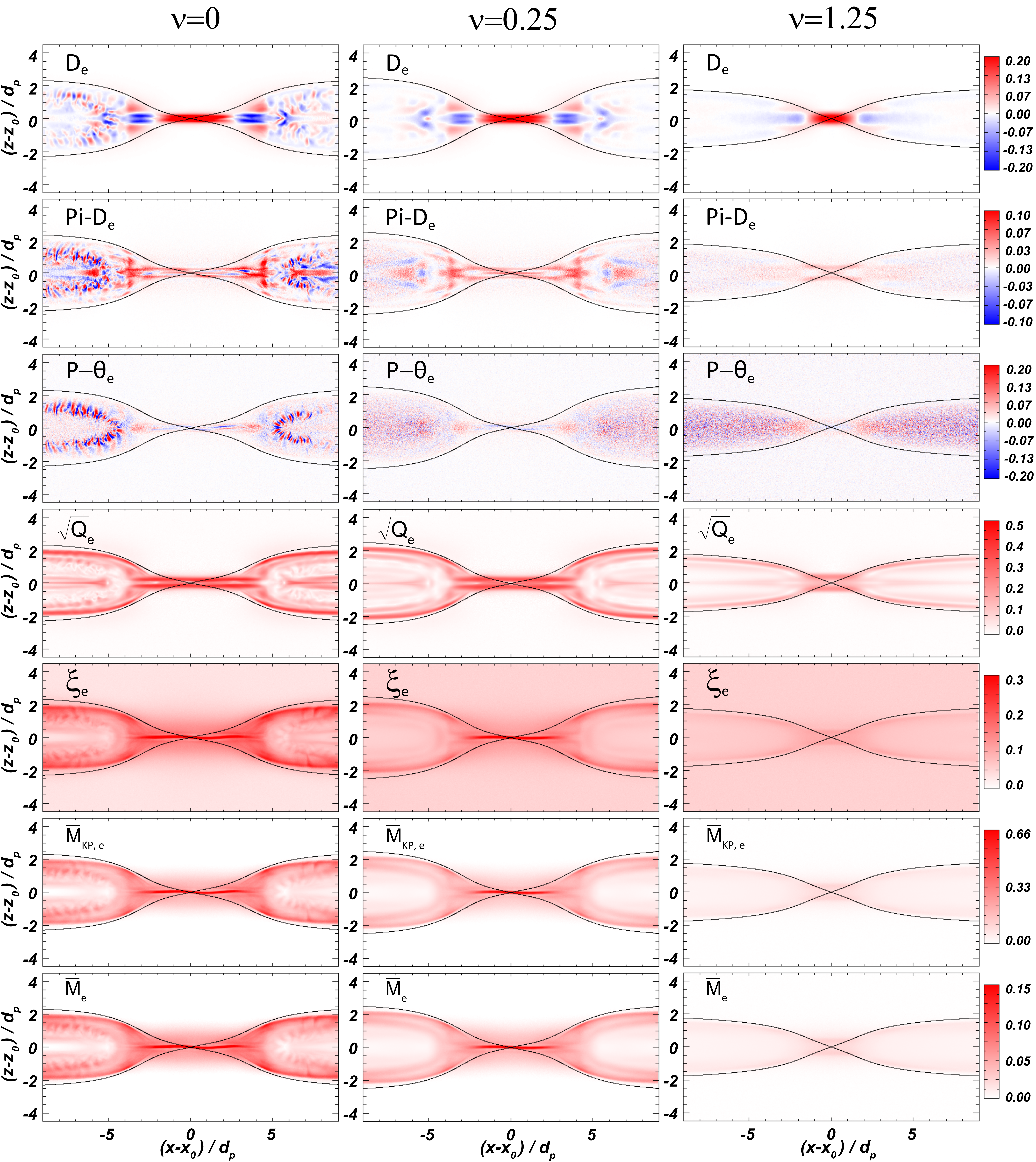}
\caption{Same as figure~\ref{fig:vpic_i}, but for electrons instead of protons.}
\label{fig:vpic_e}
\end{figure*}

\subsection{Electron dissipation proxies for collisionless reconnection}
\label{sec-elecrecon}

In analogy with protons, electron dissipation proxies are displayed in Fig.~\ref{fig:vpic_e}. One exception is that the LET parameter is not computed for electrons, as discussed in Section~\ref{sec-energymeasures}. As expected, $D_e \simeq D_p$ due to quasi-neutrality. Again, we first focus on the collisionless case in the left column.

The other energy-based parameters, the electron pressure--strain interaction terms Pi-D$_e$ and P-$\theta_e$, have highly structured patterns at much smaller scales than their proton counterparts, as expected. In the current sheet, Pi-D$_e$ and P-$\theta_e$ are both confined to the EDR, with almost no signal in the IDR. Pi-D$_e$ is positively peaked in the magnetic island close to the X-point. There are strong bands of Pi-D$_e$ near the upstream edges of the EDR, where velocity shear due to electron meandering orbits is significant. In the island, both Pi-D$_e$ and P-$\theta_e$ have strong variations in the small-scale structures discussed in the previous subsection. The intense electric fields are expanding and compressing the electron fluid as seen in P-$\theta_e$, and these fluctuations in the local velocity shear give a non-zero Pi-D$_e$. There is also coherent structure of Pi-D$_e$ as the electrons leave the EDR and develop a velocity shear as they move around the pre-existing magnetic island.

For the VDF-based proxies, all four proxies $\sqrt{Q}_e, \xi_e, \bar{M}_{\rm KP,e}$ and $\bar{{M}}_e$ are peaked in the EDR where the strong signature of $D_e$ is present. For $\sqrt{Q}_e$, it is peaked at the upstream edges of the EDR where the meandering orbits meet the upstream electrons, and is relatively smaller near the magnetic field reversal where distributions have a characteristic wedge shape \citep{Ng11}. Interestingly, the agyrotropy $\sqrt{Q}_e$ is only non-zero in the EDR, while the non-Maxwellianities $\xi_e, \bar{M}_{\rm KP,e}$ and $\bar{{M}}_e$ are non-zero in both the EDR and IDR. The reason for this is that the electrons upstream in the IDR are trapped \citep{Egedal05}, and it has been shown that they produce gyrotropic distributions elongated in the parallel direction \citep{Egedal08,Egedal13}. Consequently, the agyrotropy in the IDR is zero, while the non-Maxwellianity is non-zero, as is seen in the simulation results.

All four VDF-based proxies also show strong signatures close to the separatrices, where complicated distributions at the boundaries between upstream plasma and the magnetic island occur. This is a key distinction between these proxies and the energy-based proxies, which are not peaked near the separatrices. This signature suggests that the energy conversion in the island and exhaust is not taking place near the edges of the islands, but more towards the core as the bent field lines straighten. The VDF-based proxies also display non-zero signals at the small-scale structures in the exhaust.

\subsection{Collisional effects on proton and electron dissipation proxies} \label{sec-collisions}

We now turn to the effect of interspecies and intraspecies collisions on the dissipation measures for both protons and electrons. To put the numerical collisionality in perspective, we compare it to two known critical collision frequencies for reconnection. Collisionless (Hall) reconnection transitions to collisional (Sweet--Parker) reconnection at a critical resistivity $\eta_c$ \citep{cassak2005catastrophe}. The initial current sheet thickness $w_0 = 0.5$ is about four times smaller than $d_p$, so it is expected that collisionless reconnection will occur for small enough resistivity. From Fig.~3 of \citet{cassak2005catastrophe}, the critical resistivity is $\eta_c c^2 / 4 \pi c_A d_p \simeq 0.2$, which in normalized units for this study is $\eta_c \simeq 1$.  Then, the critical collision frequency is $\nu_c = \eta_c n_e e^2 / m_e \simeq 5$. Therefore, for $\nu = 0.25$ and 1.25, as is used here in the {\sc vpic} simulations, reconnection is expected to remain Hall-like. The time-scale for magnetic diffusion in the electron current sheet is $4 \pi d_e^2 / \eta c^2 = 1 / \nu$, so the diffusion time-scales are 4 and 0.8 for $\nu =$ 0.25 and 1.25, respectively. In comparison, the electron Alfv\'en transit time through the EDR is $2 d_e / (0.1 c_{Ae}) \simeq 2$, where $c_{Ae}$ is the electron Alfv\'en speed. Consequently, collisions are expected to have a noticeable effect in the EDR in the {\sc vpic} simulations for $\nu = 0.25$, and a significant effect for $\nu = 1.25$. A second critical collision frequency is that at which collisions affect electron trapping upstream of the EDR, which is approximately $\nu = 0.1$ \citep{le2015transition}. Thus, the trapping of electrons will be minorly affected for $\nu = 0.25$ and significantly affected for $\nu = 1.25$.  In contrast, the collisionality for the {\tt Gkeyll} simulation is very low, below both thresholds, so the evolution is essentially collisionless.

Figure~\ref{fig:vpic_i} displays the proton dissipation proxies for the $\nu=0.25$ (centre) and $\nu=1.25$ (right) {\sc vpic} simulations. Data are from $t\simeq 26$ for both $\nu = 0.25$ and $1.25$, when the magnetic energy of these simulations is nearly the same as the collisionless case at $t = 22$. The $\nu = 0.25$ case is just after the peak in reconnection rate, as for the $\nu = 0$ case. The $\nu = 1.25$ case is just before the peak in reconnection rate, which explains why the island is somewhat smaller for this case. 

The energy-based parameters should be affected by the presence of inter-species collisions due to the exchange of energy between species. For $\nu = 0.25$, collisions affect the small-scale structures that were present in the collisionless case, especially in $D_p$ and P-$\theta_p$.  However, as expected, the large-scale structure of these parameters is not greatly altered for this collisionality. For $\nu = 1.25$, however, collisions significantly alter the dissipation proxies. The signals in the magnetic islands are severely weakened, as are the signals in the EDR and IDR. Despite being weaker, the large-scale structure of the measures is largely unchanged.  

In a similar way, VDF-based parameters are qualitatively unaffected by collisional effects for $\nu=0.25$, with only weak quantitative differences. On the other hand, for $\nu=1.25$, VDF-based parameters are strongly quantitatively affected. As collisions drive distributions toward Maxwellianity, especially those with fine velocity--space structures, i.e. the non-Maxwellianity measures $\xi_p$, $\bar{M}_{\rm KP,p}$, and $\bar{{M}}_p$, are strongly decreased. The agyrotropy $\sqrt{Q}_p$ is also reduced by strong collisions, but not as drastically as the non-Maxwellianity measures, as it is less sensitive to sharp peaks in velocity space. This result again confirms that collisional dissipation acts on different characteristic time-scales depending on the scale of the velocity--space distortion in the particle VDF: finer velocity space structures produce shorter dissipation time-scales \citep{landau1936transport, rosenbluth1957fokkerplanck, balescu1960irreversible, pezzi2016collisional}. 

We analyze now the effect of collisions on electron dissipation proxies for the {\sc vpic} simulations, which bears many similarities to the effect on proton dissipation proxies. The small-scale structures in the island largely disappear even for the weaker collisionality of $\nu = 0.25$.  However, at variance with the protons, P-$\theta_e$ shows persistent small-scale structure even at large collisionality $\nu = 1.25$. The non-Maxwellianity proxies for electrons are very small for $\nu = 1.25$. This is consistent with collisions being dynamically important on the time-scale of the electron transit through the EDR; they Maxwellianize almost fully. In contrast, the EDR remains clearly visible in the non-Maxwellianity measures for $\nu = 0.25$. The agyrotropy is non-zero at the edges of the EDR for all three simulations, suggesting that the meandering orbits are sufficient to produce this signal even for the highest collisionality considered here. The trapped electrons upstream of the EDR are weaker for $\nu = 0.25$ and nearly non-existent for $\nu = 1.25$, consistent with the predictions from \citet{le2015transition}.

We finally describe the weakly collisional continuum \texttt{Gkeyll} simulation. The \texttt{Gkeyll} simulation includes only the effects of intraspecies collisions. 

\begin{figure}
\centering
\includegraphics[width=0.48\textwidth]{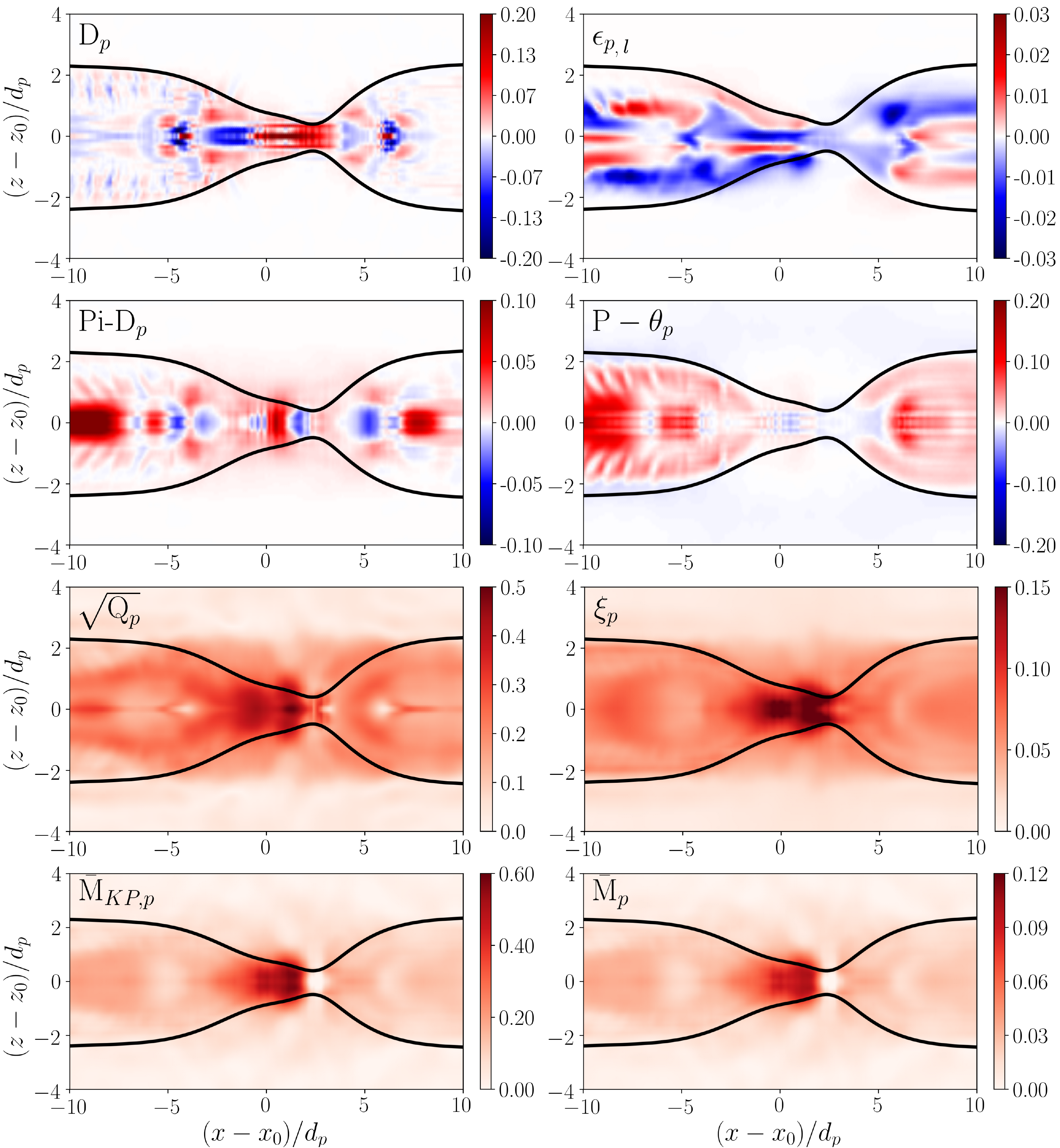}
\caption{Same dissipation proxies as plotted in Figure~\ref{fig:vpic_i}, but for the protons in the {\tt Gkeyll} reconnection simulation.}
\label{fig:proton-gkeyll}
\end{figure}

Figure~\ref{fig:proton-gkeyll} shows the proton dissipation proxies, plotted at a slightly earlier time $t\simeq 18$, but this time is after the peak of the reconnection rate which takes place at an earlier time with respect to the {\sc vpic} case owing to the smaller system size. The plots reveal that the X-line is not located exactly in the center of the domain and there is a significant left--right asymmetry. This is because the random perturbation to the initial condition seeding reconnection breaks the symmetry in the $x$ direction.

Three of the energy-based measures, $D_p$, Pi-D$_p$, and P-$\theta_p$, show broad agreement with the {\sc vpic} results in Figure~\ref{fig:vpic_i}. The \texttt{Gkeyll} simulation has a more structured Pi-D$_p$ near the X-line due to a secondary island near $z=z_0$ and $x-x_0\simeq[1,2]$, which develops shortly after the peak of the reconnection rate. The secondary island has a bipolar structure in Pi-D$_p$ analogous to the bipolar structure downstream in a dipolarization front in the {\sc vpic} simulation, visible at $z=z_0$ and $x-x_0\simeq[-9,-4]$ in Figure \ref{fig:vpic_i}. The strongest difference between the two data sets is observed in the LET diagnostic $\epsilon_p$, which displays broader features in the \texttt{Gkeyll} simulation. This is likely due to the coarser configuration-space resolution. 

For the VDF-based diagnostics, the \texttt{Gkeyll} simulation gives qualitatively and quantitatively similar results in the exhaust, especially the inner shell of the magnetic island. However, small differences arise in these diagnostics due to the secondary island in the \texttt{Gkeyll} simulation. Indeed, this structure generates intense deviations from the thermal equilibrium due to the mixing and rapid rotation of protons trying to align with the local magnetic field. Inside the proton scale island, we observe strong deformation of the proton VDF which manifests as an intense agyrotropy $\sqrt{\textrm{Q}_p}$ and non-Maxwellianity $\xi_p$, $\bar{\textrm{M}}_{KP,p}$ and $\bar{\textrm{M}}_{p}$. 

\begin{figure}
\centering
\includegraphics[width=0.48\textwidth]{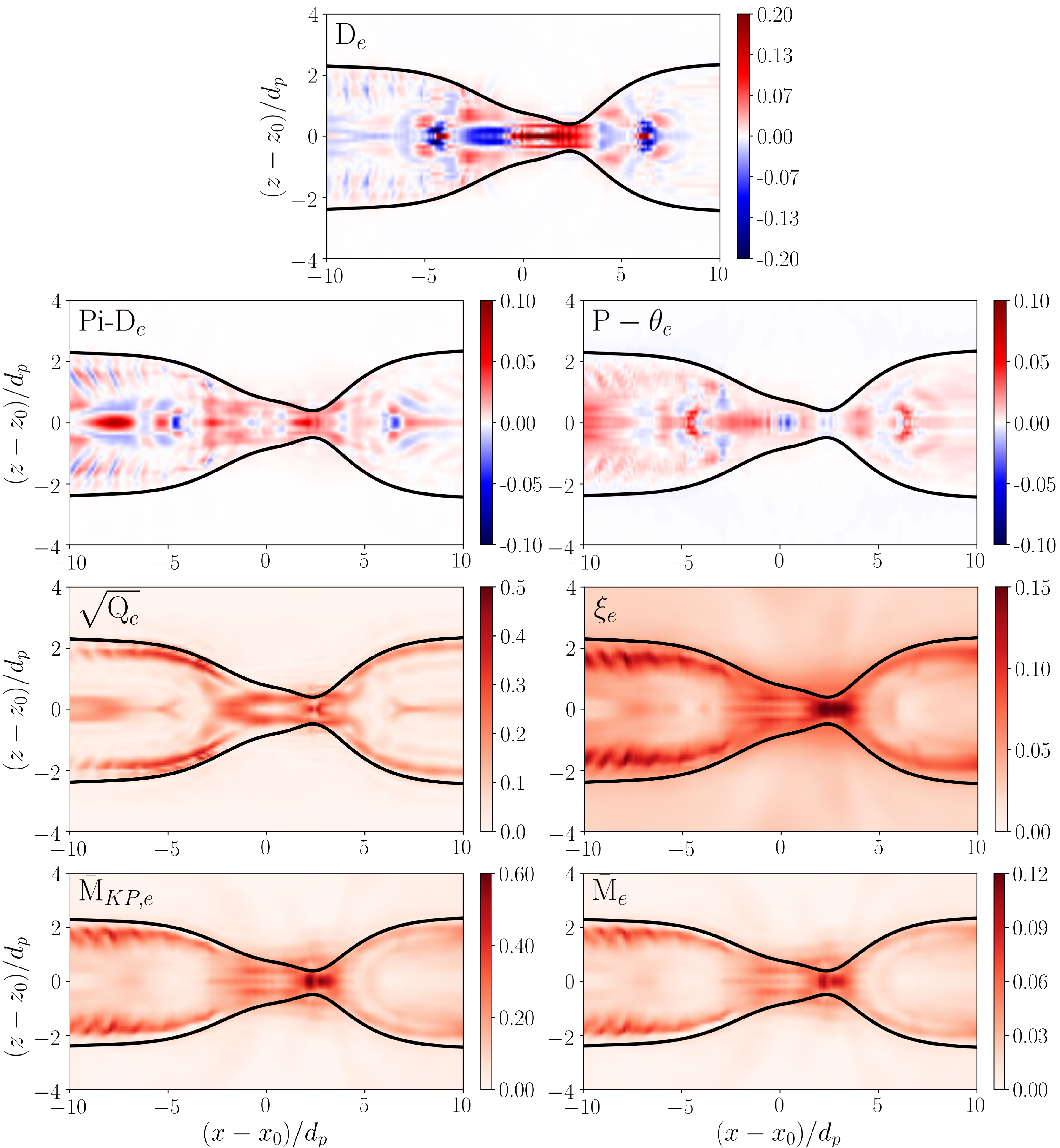}
\caption{Same dissipation proxies as plotted in Figure~\ref{fig:vpic_e}, but for the electrons in the {\tt Gkeyll} reconnection simulation.}
\label{fig:elc-gkeyll}
\end{figure}

Turning to electron dissipation proxies, displayed in Figure \ref{fig:elc-gkeyll}, the \texttt{Gkeyll} simulation results display a good agreement with the {\sc vpic} data. In fact, the overall structure in energy-based diagnostics such as $D_e$ and Pi-D$_e$ and distribution function-based diagnostics such as $\sqrt{\textrm{Q}_e}$ and $\xi_e$, agrees better for electrons than for protons. For example, $D_e$ and Pi-D$_e$ are positive in the electron diffusion region, and we observe enhancement of all the distribution function-based diagnostics near the separatrices. This better agreement can be linked to the secondary island not having as dramatic an impact on the electron dynamics in the magnetic island. 

The comparison of this wide array of diagnostics from these two different codes in different regimes, from collisionless {\sc vpic} to weakly collisional \texttt{Gkeyll} to collisional {\sc vpic}, reveals the diversity of information content each diagnostic contains. In many cases, we observe little qualitative difference between the energy-based diagnostics from different simulations while VDF-based diagnostics are more sensitive both qualitatively and quantitatively to the strength of collisions and subtle differences in the underlying kinetic evolution of the reconnection process, such as the secondary island which forms in the \texttt{Gkeyll} simulation. 

\section{Numerical results: Dissipation measures in turbulent plasmas}
\label{sect:turb}

\begin{figure*}
\begin{minipage}{0.48\textwidth}
   \centering
\includegraphics[width=0.9\textwidth]{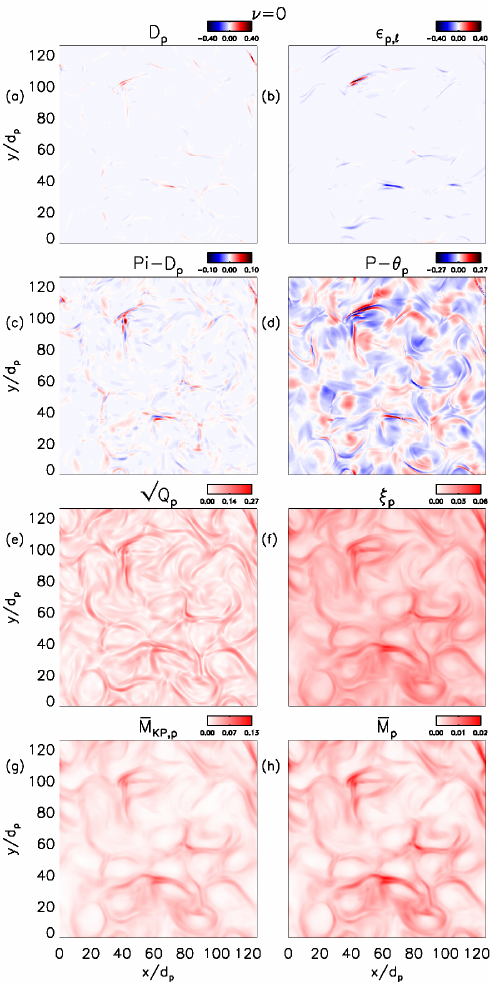}
  \end{minipage}
 \begin{minipage}{0.48\textwidth}
 \centering
   \includegraphics[width=0.9\textwidth]{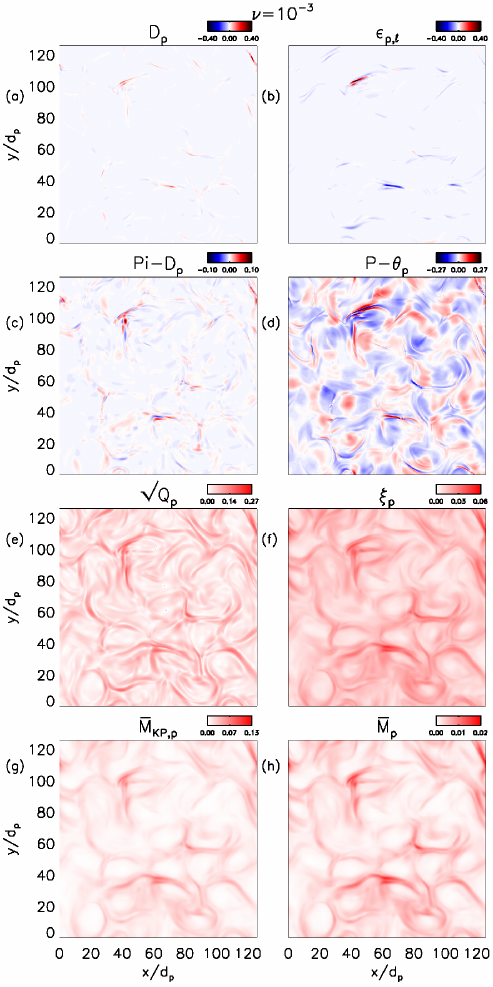}
 \end{minipage}
\caption{Various dissipation surrogates evaluated at the time of maximum turbulent activity in HVM simulations, $t=t^*$. The two columns at left refer to the collisionless case, and the two at the right are for the weakly collisional case. Panels from (a) to (h) display $D_p$, $\epsilon_{p,\ell}$ with $\ell\simeq d_p$, Pi-D$_p$,  P-$\theta_p$, $\sqrt{Q_p}$, $\xi_p$, $\bar{M}_{\rm KP,p}$, and $\bar{M}_p$, respectively. }
\label{fig:HVM_t1}
\end{figure*}

We here describe the structure of dissipation proxies in plasma turbulence at kinetic scales. Since the HVM code using the hybrid model neglects electrons, we only treat proton parameters. In these simulations, energy injected at large scales generates a cascade towards smaller scales. The time corresponding to the most intense turbulent activity is $t=t^*=30$, in which a turbulent state characterized by an intermittent pattern of current sheets that border magnetic islands and vortices \citep{servidio2015kinetic, wan2015intermittent} is reached. We show simulation results at $t=t^*$ in Fig.~\ref{fig:HVM_t1}, with the output of the collisionless (left) and collisional (right) simulation. Panels from (a) to (h) display $D_p$, $\epsilon_{p,\ell}$ with $\ell\simeq d_p$, Pi-D$_p$, P-$\theta_p$, $\sqrt{Q_p}$, $\epsilon_p$, $\bar{M}_{\rm KP,p}$, and $\bar{M}_p$, respectively. 

We begin by analysing the energy-based parameters. $D_p$, LET $\epsilon_{p,\ell}$, and the Pi-D$_p$ term are all peaked close to the most intense current sheets. This confirms that current sheets are the sites with the most intense local energy conversion and dissipation. $D_p$ has a preferred sign, being positive in most of the regions of highest energy conversion. This implies there is a net conversion of energy from the electromagnetic fields to protons. Similarly, the predominantly negative LET supports the standard picture of a direct global energy cascade towards small scales. The regions of larger energy transfer are generally located near the current sheets, and some complexity in the fine local details of the transfer can be observed. A detailed analysis of each term of the right-hand side of Eq. (\ref{let}) reveals that the total energy (kinetic plus magnetic) available to be transported by the longitudinal component $\Delta u_{p,\ell}$ is the main contributor to the LET parameter in this HVM simulation. In this case, it is also confirmed (not shown) that the global (Yaglom--Hall) law shows a linear scaling in the interval roughly corresponding to the MHD-turbulence range ($2 d_p \lesssim \ell \lesssim 10 d_p$), as also recently reported in similar simulation setups by \citet{sorriso2018local} and \citet{vasconez2020local}.

As displayed in previous HVM \citep{pezzi2019energy} and in PIC simulations \citep{yang2018scale}, Pi-D$_p$ is highly structured, having both positive and negative regions close to intense current sheets. Conversely, P-$\theta_p$ is larger than its Pi-D$_p$ counterpart and has significant contributions both at the current sheets and in magnetic islands since it is related to large-scale plasma compression (red) and rarefaction (blue). The most intense regions of P$-\theta_p$ are at current sheets, reflecting the rapid collision or separation of large-scale magnetic islands.

\begin{figure*}
\centering
\begin{minipage}{0.35\textwidth}
   \centering
\includegraphics[width=1\textwidth]{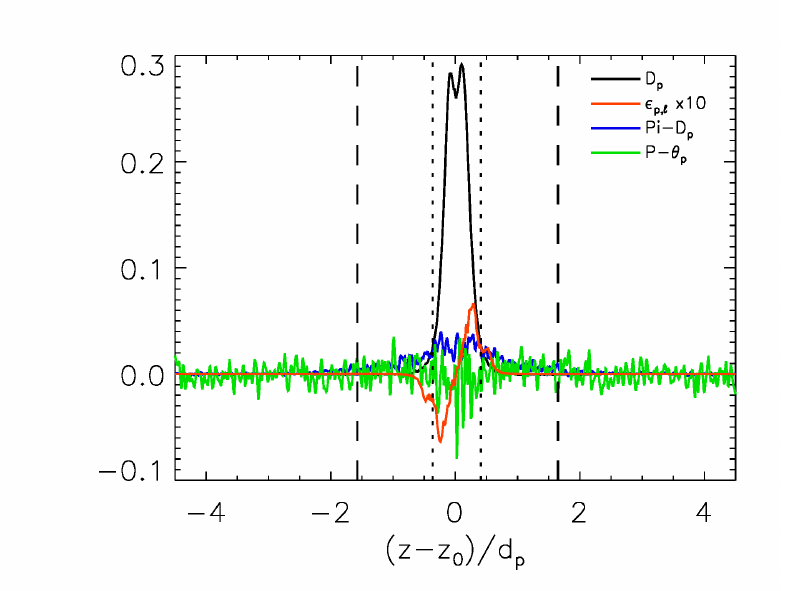}
  \end{minipage}
 \begin{minipage}{0.35\textwidth}
 \centering
   \includegraphics[width=1\textwidth]{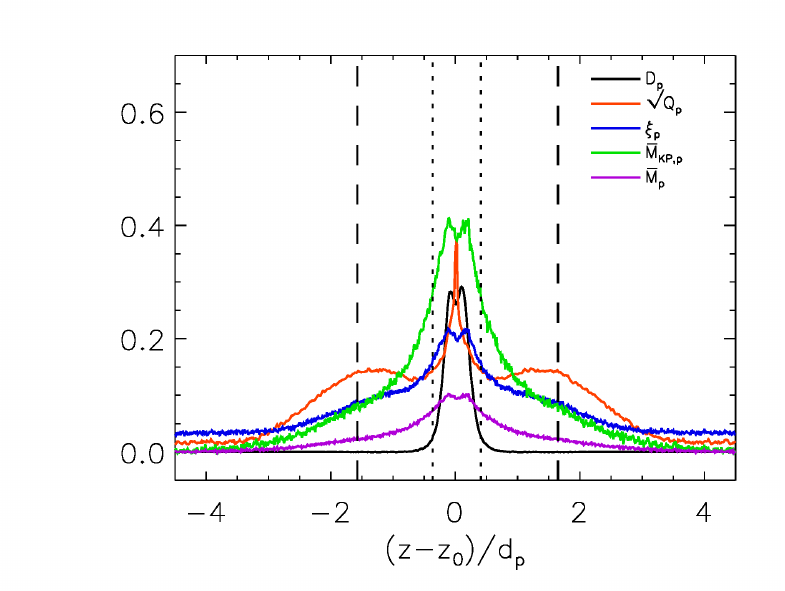}
 \end{minipage}
 \begin{minipage}{0.35\textwidth}
   \centering
\includegraphics[width=1\textwidth]{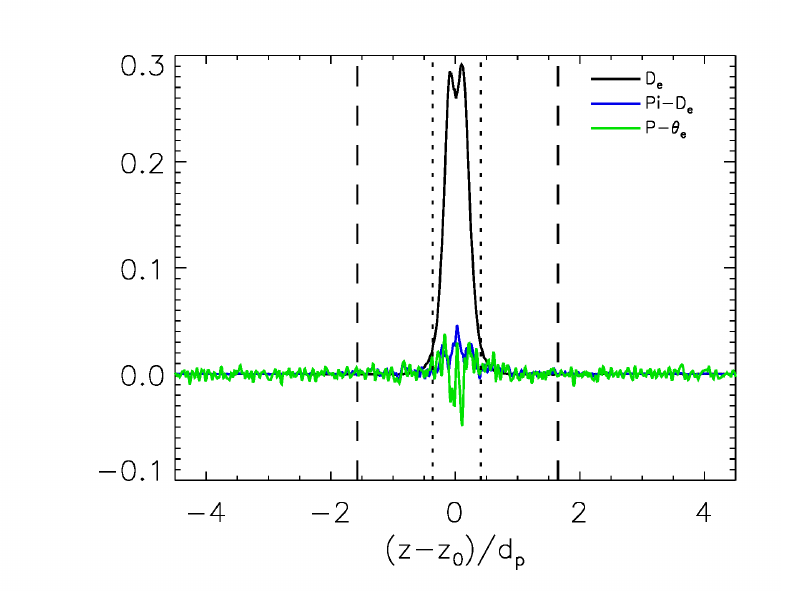}
  \end{minipage}
 \begin{minipage}{0.35\textwidth}
 \centering
   \includegraphics[width=1\textwidth]{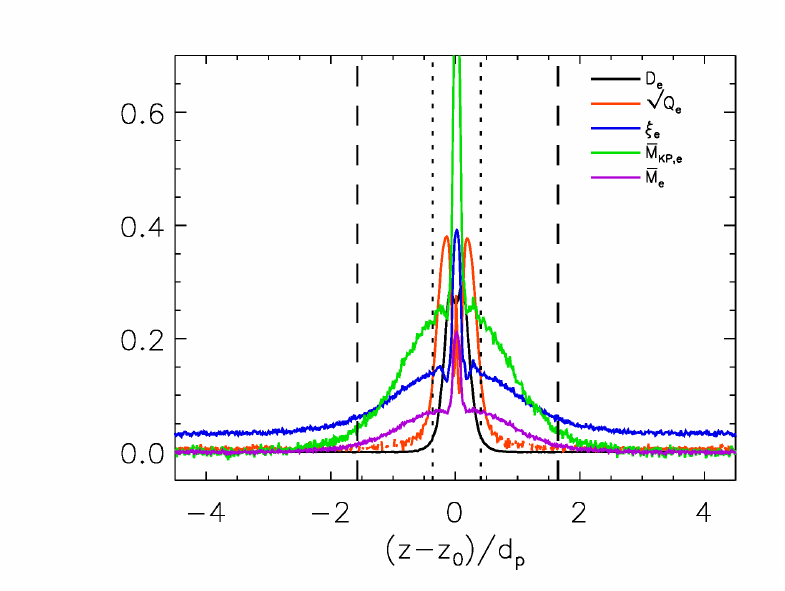}
 \end{minipage}
\caption{1D cuts vertically through the $x$-line showing the various dissipation surrogates in the reconnecting current sheet for the {\sc vpic} collisionless reconnection simulation. The top row refers to proton proxies, displaying energy-based parameters (left, with $\epsilon_{p,\ell}$ scaled by a factor of 10 to make it more visible) and VDF-based parameters (right). In the right-hand panel, the Zenitani measure $D_p$ is showed as a reference. The bottom row is analogous for the electron proxies.}
\label{fig:VPIC_cut_x0}
\end{figure*}

The four VDF-based proxies bear many similarities. They are highly structured, with local peaks close to current sheets. As with the reconnection simulations, there are also differences between these measures. It is more common to get appreciable values of $\sqrt{Q}_p$ than other VDF-based measures, especially $\bar{M}_{KP,p}$ and $\bar{M}_{p}$. This indicates that the agyrotropy provides an overall picture of the presence of large-scale kinetic effects in the VDFs (namely the $2^{\rm nd}$ order VDF moment), while fine-scale structures in the VDF are larger contributors to non-Maxwellianity measures.

The collisionless and weakly collisional simulations do not reveal significant differences in the energy-based parameters for the collision frequency in use. The inclusion of proton--proton collisional effects does not affect the statistical characteristics of turbulence at the proton scale. This can be explained since, at variance with interspecies collisional effects, intraspecies collisions do not generate a resistivity-like term which directly affects the electric field and hence fluid quantities. On the other hand, VDF-based parameters are dissipated by collisions, signature of the collisional thermalization. Since these parameters are sensitive to the presence of out-of-equilibrium structure in the proton VDF, they are affected by intraspecies collisional effects. The effect of collisions is less visible in $\sqrt{Q_p}$ since collisions preferentially dissipate fine velocity--space structures \citep{pezzi2016collisional}, which contribute less to $\sqrt{Q_p}$. The effect of collisions is visible in $\xi_p$ and in the entropy density-based non-Maxwellianity proxies. The maximum values of these three parameters are smaller by about $10\%$ in the weakly collisional simulation than the collisionless simulation. Similarly, average values are reduced by about $20\%$ for $\xi_p$ and $30\%$ for $\bar{M}_{\rm KP,p}$ and $\bar{M}_p$ when including collisions. The effect of collisions becomes more significant at later times \citep[not shown here, see Fig. 4 of ][]{pezzi2019protonproton}. Indeed, at the final time of the simulation, maxima of $\xi_p$ and of the entropy density-based proxies are, respectively, about $20$ and $30\%$ smaller when collisional effects are considered, while their average values are reduced by $40$ and $60\%$, respectively. The different dissipation (in terms of both maximum and averaged values) of $\xi_p$ and the entropy density-based proxies suggests that the latter react more to collisional effects than $\xi_p$ since entropy-based proxies are dissipated more efficiently via collisions.

\section{One-dimensional profiles of dissipation proxies}
\label{sect:1dcuts}

We conclude by presenting 1D cuts of the dissipation surrogates adopted in this work across a typical current structure. We include both the reconnecting current sheet and a typical current sheet observed in the turbulent HVM simulation. We focus on the collisionless simulation, since our aim is mainly discussing how these parameters look when crossing a particular structure. Moreover, for the magnetic reconnection simulations, we show cuts for only the {\sc vpic} runs, since \texttt{Gkeyll} and {\sc vpic} provide similar results. These plots help reveal comparative proxy structure near to the current sheets, and may also be useful for future comparison with {\it in situ} spacecraft observations, e.g., for the Pi-D measure \citep{chasapis2018energy}. For applications of these results to observations or laboratory experiments, it is important to emphasize that the results are undoubtedly sensitive to the plasma parameters and to the reduced 2D physical-space dimensionality \citep{howes2015inherently,li2015dissipation}. Note also that multispacecraft observations are necessary to compute the spatial derivatives needed to calculate some of these diagnostics (e.g. Pi-D measure).

\begin{figure*}
\centering
\begin{minipage}{0.35\textwidth}
   \centering
\includegraphics[width=1\textwidth]{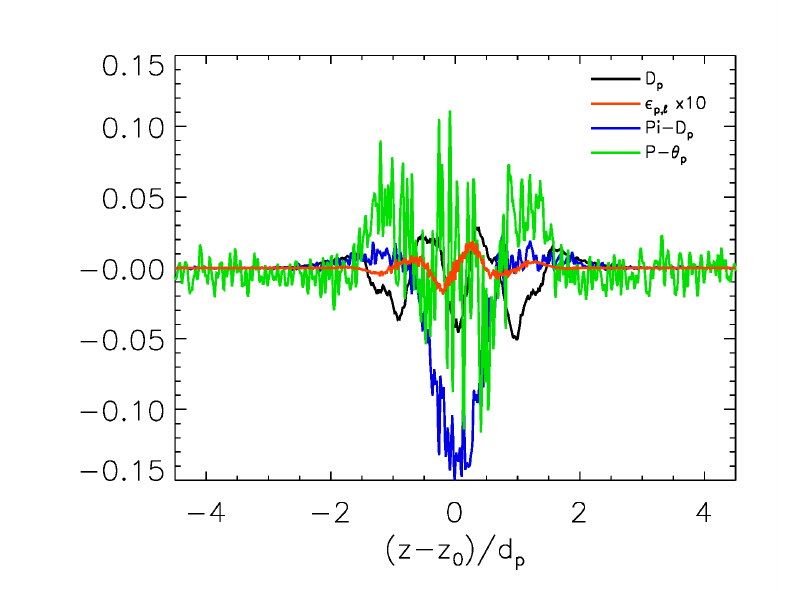}
  \end{minipage}
 \begin{minipage}{0.35\textwidth}
 \centering
   \includegraphics[width=1\textwidth]{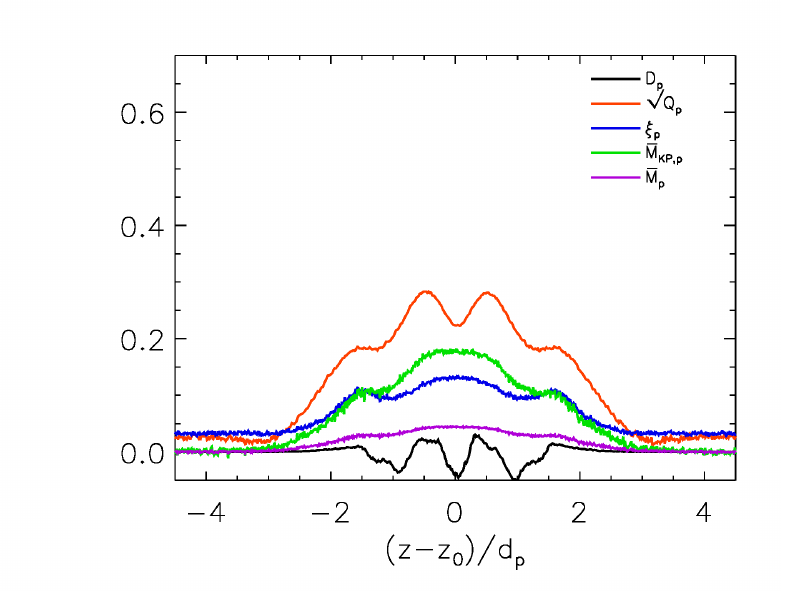}
 \end{minipage}
  \hfill
 \begin{minipage}{0.35\textwidth}
   \centering
\includegraphics[width=1\textwidth]{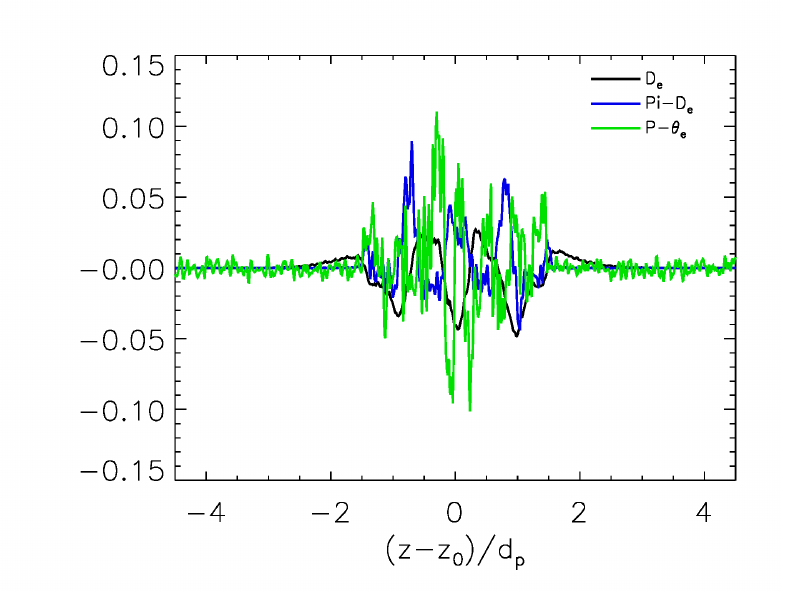}
  \end{minipage}
 \begin{minipage}{0.35\textwidth}
 \centering
   \includegraphics[width=1\textwidth]{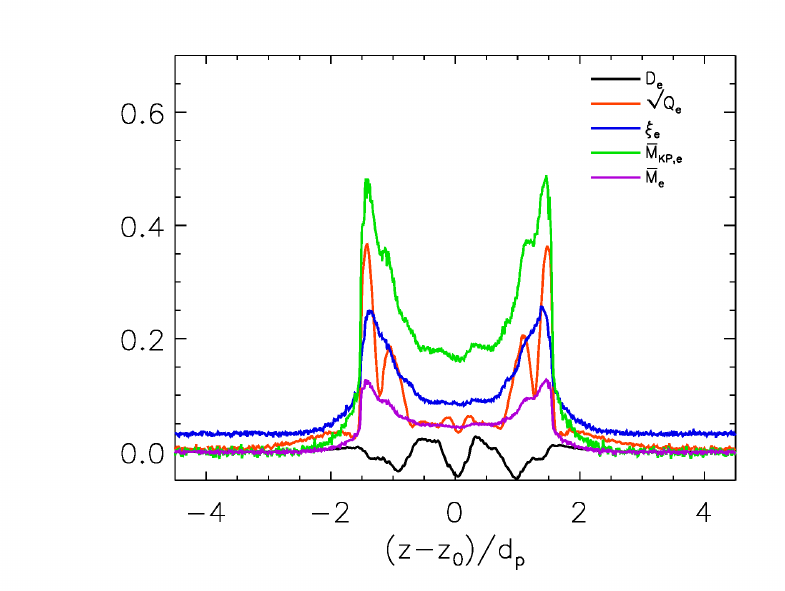}
 \end{minipage}
\caption{Same as figure~\ref{fig:VPIC_cut_x0}, but at $x-x_0=5 d_p$.}
\label{fig:VPIC_cut_x5}
\end{figure*}

Figure \ref{fig:VPIC_cut_x0} shows the dissipation proxies along a cut in the reconnection simulation through the X-point in $z$. The vertical dotted and dashed lines mark the upstream edges of the EDR and IDR, respectively. These edges are defined by the location at which the electron and ion out-of-plane currents are $20\%$ of their maximum \citep{shay2001alfvenic}. The upper panels show the results for the protons, and the lower panels are for electrons, with energy-based measures in the left plots and VDF-based measures on the right. For the energy-based measures, $D_e \simeq D_p$ shows a clear peak in the EDR. The LET parameter $\epsilon_{p,\ell}$, scaled by a factor of 10 to make it easier to see, displays the negative-positive double-peaked structure in the EDR seen in Figure \ref{fig:vpic_i}, and is negligible elsewhere. The sign of $\epsilon_{p,\ell}$ is the same sign as $B_x$ for this current sheet. The electron and ion Pi-D and P-$\theta$ show moderately intensified signals within the EDR with tails extending into the IDR. Compared to the Zenitani measure, both P-$\theta$ and Pi-D display rapidly fluctuating patterns but only Pi-D is positive definite within the IDR, i.e. the same sign of $D_p$.

For the VDF-based measures, the agyrotropies $\sqrt{Q_e}$ and $\sqrt{Q_p}$ reveal a double-peak structure the diffusion region of each species correlated with the size of its diffusion region. As discussed in Section~\ref{sec-elecrecon}, this shape is caused by the meandering motions of the particles traversing their diffusion regions at their gyroscale and is therefore a characteristic shape of $\sqrt{Q}$ in the diffusion region of antiparallel reconnection. For both electrons and ions, the non-Maxwellianity proxies $\xi$, $\bar{M}_{KP}$, and $\bar{M}$, show intensified signals over the IDR with relatively strong peaks in the EDR. As discussed in Section \ref{sect:MR}, the electron non-Maxwellianities not only show the peaks in the EDR but also broad intensified signals over the IDR. As analysed in \citet{liang2020kinetic}, the broad intensified signals are due to the gyrotropic distributions created by trapped electrons \citep{Egedal05}. This non-zero signal is coincident with negligible signal in the electron energy-based proxies $D_e$, Pi-D$_e$ and P-$\theta_e$.

Figure \ref{fig:VPIC_cut_x5} similarly displays the  dissipation proxies along a vertical cut through the reconnection simulation at $x-x_0=5$, i.e. through the reconnection exhaust. The format is the same as Figure \ref{fig:VPIC_cut_x0}. For this cut, the exhaust is between the separatrices at $z-z_0 \simeq \pm 1.5$. For both electrons and ions, all energy-based diagnostics, $D_e = D_p$, $\epsilon_p$, Pi-D and P-$\theta$ show intensified, yet noisy, signals. The signals include positive and negative values due to complex flows in this region, although for this cut the Pi-D$_p$ measure has a sizeable negative value. The VDF-based proxies for both electrons and ions are intensified in the exhaust, as well. The electron VDF-based proxies have peaks near the separatrices, due to the counter-streaming electron flows \citep{hesse2018role,liang2020kinetic}. The VDF-based proxies are slightly broader than the exhaust, which results from the finite Larmor radius (FLR) effects near the separatrices as shown by the non-gyrotropy $\sqrt{Q_p}$. Although FLR effects are not seen much in the energy-based diagnostics, they are picked up by the VDF-based diagnostics $\xi_p$, $\bar{M}_{KP,p}$ and $\bar{M}_p$. 

\begin{figure}
\centering
\begin{minipage}{0.35\textwidth}
   \centering
\includegraphics[width=1\textwidth]{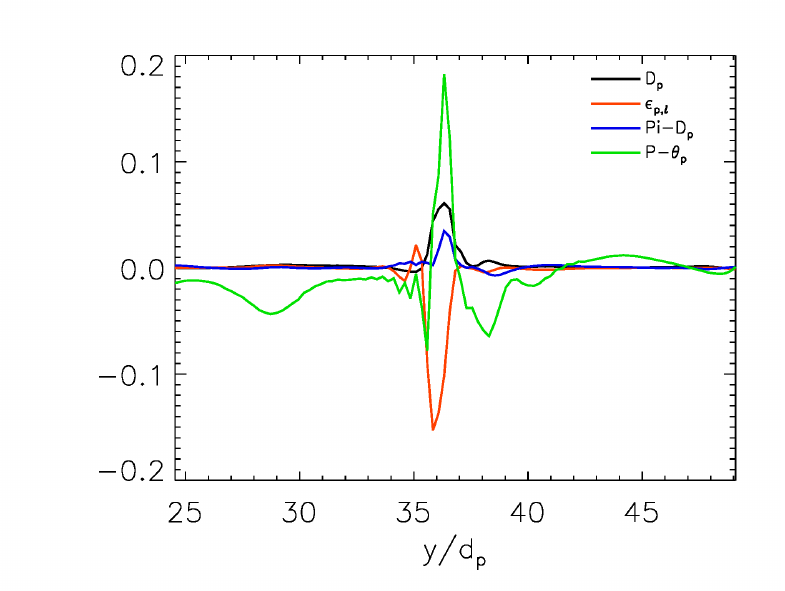}
  \end{minipage}
 \begin{minipage}{0.35\textwidth}
 \centering
\includegraphics[width=1\textwidth]{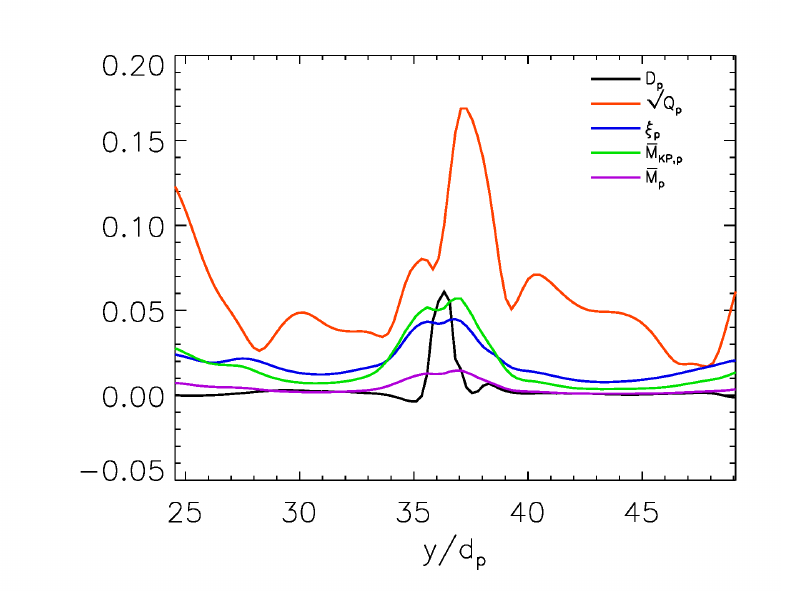}
 \end{minipage}
\caption{(Color online) Same as figure~\ref{fig:VPIC_cut_x0}, but for the HVM collisionless turbulence simulation. The cut is taken at $x=68.7 \ d_p$. }
\label{fig:HVM_cut}
\end{figure}

Finally, we show in Fig. \ref{fig:HVM_cut} the dissipation proxies for the collisionless HVM turbulence simulation along a 1D cut close to the current sheet near $(x,y) = (69,36)$. Similar results are obtained in the collisional simulation. All the energy-based parameters (top panel of Fig.~\ref{fig:HVM_cut}) are peaked within the same region, i.e. $y\simeq 36$. The VDF-based proxies (bottom panel of Fig.~\ref{fig:HVM_cut}) also reveal clear maxima near the current structure. However, the maxima are somewhat broader in width ($\sim 2-5$) than the energy-based parameters ($\sim 1$). Moreover, the agyrotropy parameter $\sqrt{Q_p}$ shows large-scale fluctuations quite far from the current structure. The broader distribution observed for the VDF-based parameters suggests that energy transfer is localized close to the coherent structures, but can affect the particle distribution function in a larger region around these structures. 

The 1D cuts of the dissipation proxies for both the reconnection and turbulence simulations support the idea that peaks of dissipation measures in a reconnecting current sheet or a turbulent environment characteristically occur in coincident spatial regions, but not necessarily at the same exact spatial position. This is consistent with the notion of {\it regional} correlations, suggested by \citet{yang2018scale,matthaeus2020pathways}. The selected structure has more of a resemblance to the 1D cut of the reconnection simulation at the X-point than at the reconnection exhaust. However, there are qualitative and quantitative differences. The Zenitani measure and the diagnostics based on the pressure--strain interaction (Pi-D and P-$\theta$) have the same dimensions (see Sect. \ref{sect:nummodel}), hence it is reasonable to compare their magnitude. In this respect, we notice that the Zenitani measure is most prominent in the reconnection simulation, while a positive P-$\theta$ is the strongest in the turbulence simulation. This suggests that, when simulating a single reconnecting current sheet, compressible effects are less significant. On the other hand, when reconnection occurs in a turbulent environment, where magnetic islands can merge with each other, compression is much more significant. The LET parameter, which has different dimensions than the other three energy-based surrogates, oscillates in the reconnection simulation, but is strongly negative in the turbulence one. Qualitative agreement between the structures is found for the VDF-based surrogates, that are all adimensional, although some features in the reconnection simulation (e.g. double peaks in the agyrotropy) are not present in the turbulence simulation perhaps because of the different numerical resolution. Furthermore, for the cut through the current sheet in turbulence, the parameters depend on time due to the turbulent interaction of larger scale magnetic eddies, so wide variation in parameters across different current sheets and at different times is expected.

\section{Discussion and conclusions}
\label{sect:concl}

As a result of the weak collisionality of many space and astrophysical plasmas, several physical processes can locally contribute to the dissipation of energy and the heating of the plasma. During magnetic reconnection, magnetic energy is efficiently converted to directed plasma flows, thermal energy, and energetic particles \citep{burch2016electron,torbert2018electronscale}. During turbulence in a magnetized plasma, the cascade provides the global amount of energy needed for fluctuations to be dissipated at small scales \citep{marino2008heating}. However, which dissipative mechanisms are dominant and under which conditions is still not sufficiently understood \citep{vaivads2016turbulence}. Therefore, a multiplicity of dissipation surrogates have been adopted in the literature to identify potential sites of dissipation. Owing to the strongly non-linear dynamics and the importance of physics at kinetic scales in such systems, numerical simulations of the Vlasov--Maxwell system (or, including collisions, the Boltzmann--Maxwell system) are the decisive tool to address the long-standing issue of plasma heating and energy dissipation in magnetized plasmas, e.g. \citet{parashar2009kinetic}. For example, the ``turbulence dissipation challenge''  \citep{parashar2015turbulent} motivated numerical work to compare different algorithms on the solution of similar problems to assess the nature of the dissipation in magnetized turbulence (e.g. \citet{pezzi2017colliding, perrone2018fluid, gonzalez2019turbulent}). 

In the same spirit, we have conducted a survey of a number of dissipation surrogates with three different kinetic plasma codes --- the fully kinetic particle-in-cell {\sc vpic}, the fully kinetic Eulerian Vlasov--Maxwell \texttt{Gkeyll}, and the Eulerian Hybrid Vlasov--Maxwell codes --- to perform numerical simulations of two important physical phenomena in plasma physics. The first class investigates reconnection in an isolated current sheet, and the second concerns plasma turbulence at kinetic scales. We have calculated and compared eight distinct dissipation proxies, delineated in Section \ref{sect:dissmeas}. For the sake of clarity, we have categorized them in terms of (i) energy-based parameters, whose definition describes energy transfer and conversion; and (ii) VDF-based parameters, that are directly related to kinetic signatures in the particle VDF. Energy-based parameters considered here are the power density by electromagnetic fields on charged particles \citep{zenitani2011new}, the pressure--strain interaction \citep{yang2017energyPOP}, and a local proxy of the turbulent energy transfer \citep{sorriso2018local}. The VDF-based parameters are the local pressure agyrotropy \citep{swisdak2016quantifying} and three measures describing how different a local distribution function is from being Maxwellian \citep{kaufmann2009boltzmann,greco2012inhomogeneous,liang2019decomposition}. 

Our findings are that each of the studied measures is non-zero in key settings in reconnection and turbulence, including the reconnection diffusion region and magnetic islands, and the intermittent magnetic shear regions bordering magnetic eddies in a turbulent system. The region that each proxy is strongest highlights a potentially different aspect of the physics taking place, as is described in detail in Sections \ref{sect:MR} and \ref{sect:turb}. It is intended that the discussion therein will contribute to the assessment of dissipation and energy conversion in satellite and plasma laboratory experiment measurements. We here remark on the importance of the VDF-based diagnostics, which reveal further details about the underlying dynamics of the plasma compared to the energy-based diagnostics. Indeed, the energy-based diagnostics give similar results for similar simulations in the magnetic reconnection setup (i.e. {\sc vpic} and \texttt{Gkeyll} runs), as we expect since these simulations which show little difference in many of the measures of the plasma response to magnetic reconnection. However, the additional level of detail provided by VDF-based diagnostics allows us to more carefully ascertain the kinetic response of the plasma from small differences which arise naturally between two different simulation codes, and the added robustness of these diagnostics is likely to be especially useful when analysing more general simulations and observations of real plasma systems such as the solar wind. In contrast, the energy-based diagnostics are directly related to the turbulence cascade process that connects the energy containing scales to the kinetic range of scales where the velocity space structure becomes increasingly important. Therefore both type of diagnostics are helpful to form a complete picture of the dynamics leading to dissipation. 

The spatial locations where each proxy has a local maximum can be different between the different proxies. For example, in the magnetic reconnection simulations, the Zenitani measure is peaked at the X-point while the pressure--strain interaction terms are peaked inside the magnetic islands. This confirms a suggestion that both energy-based and VDF-based parameters display a regional correlation: structures identified by these parameters often occur in similar regions, although not necessarily exhibiting a point-to-point correlation. This underscores a key conclusion of this work: that no one single measure is universally the best measure for dissipation. Rather, employing a number of proxies is likely best for assessing the dissipation and energy conversion in a plasma system.

We also consider the role of collisions by including their dynamical effect {\it ab initio} in the numerical simulations for both reconnection and turbulence. When including only weak intraspecies (proton--proton) collisions in the turbulence simulations, the VDF-based proxies decreased while the energy-based proxies were largely unchanged. This behavior is expected since such collisions drive particle VDFs towards Maxwellian distributions, but they do not produce a net energy transfer between different species although they indirectly modify the energy equations by isotropizing the pressure tensor \citep{DelSartoEA16}. On the other hand, both energy- and VDF-based proxies are influenced by interspecies and intraspecies (electron--proton) collisions in simulations of magnetic reconnection. In particular, small-scale structures are dissipated by collisions, and the distributions in the diffusion region are less non-Maxwellian and thermalize more rapidly than in the collisionless case.

Interestingly, we find the peaks of dissipation proxies in the systems with collisions are generally weaker than in the collisionless system. VDF-based dissipation proxies show this behavior for all the simulations reported in the manuscript. These surrogates are local in position space and provide only a local measure of the complexity of the distribution function. Energy-based measures also become weaker in the collisional runs of magnetic reconnection, which include interspecies collisional effects. This behavior is an effect of the work of collisions that, since the beginning of the numerical simulation, have produced dissipation. In particular, intraspecies collisions dissipate VDF complexity, while interspecies collisions affect both the VDF and the energy. Therefore, by looking at these parameters at a fixed time snapshot, their peaks are weaker in the collisional runs with respect to the collisionless case. Further, it also clearly illustrates that the energy-based measures contain information about both collisional and collisionless processes, implying that they are not yet capable of distinguishing reversible from irreversible processes. 

Although defining the concept of dissipation in a collisionless plasma is rather complicated, analysing dissipation surrogates in collisionless simulations and comparing the evolution of these proxies in collisionless and collisional runs, as done in this study, is meaningful. Such a comparison indeed addresses the key question: are collisions preferentially active in regions where the dissipation surrogates evaluated in absence of collisions are intense, i.e. where other dynamical characteristics (e.g. pressure--strain interactions, non-Maxwellian structures) are significant? Our analysis may indirectly suggest a positive answer to this question. The regional correlation of energy-based and VDF-based diagnostics indicate that non-Maxwellian structures in the particle VDF are expected close to regions where energy-based dissipation surrogates are also strong. Moreover, as shown in previous studies \citep{landau1936transport, rosenbluth1957fokkerplanck,balescu1960irreversible,schekochihin2009astrophysical,pezzi2016collisional,pezzi2019protonproton}, collisions rapidly dissipate strong velocity--space disturbances in the particle VDF. Hence, we suggest that regions identified by dissipation surrogates in the collisionless simulations are regions where collisions, if present, preferentially smooth out non-Maxwellian features in the particle VDF, thus producing irreversible dissipation.

The present study is not intended to be the final word on this topic, as there are many avenues for future work. The diagnostics considered here do not discriminate the underlying process which may lead to energy dissipation or conversion, e.g. Landau damping, cyclotron damping, phase-mixing, and stochastic heating \citep{ChandranEA10,li2015dissipation}. In this perspective, a different approach, based on the {\it field-particle correlation}, has been recently adopted to identify the presence of particular signatures in the particle VDF \citep{klein2016measuring,klein2017diagnosing,chen2019evidence,klein2020diagnosing}. This method has the advantage of diagnosing energization directly in velocity space while the others here adopted involve an integration over the full velocity space. The field-particle correlation is also local in physical space and does not require spatial gradients, which would require multiple spacecraft for {\it in situ} observations. Incorporating the field--particle correlation gives a visual way to identify energy transfer between particles and fields, and would be interesting to compare with the other proxies. Indeed, the diagnostics considered in this work are extremely useful to characterize potential sites of intermittent dissipation, e.g. in structures close to intense current sheets. On the other hand, methods such as the field--particle correlation identifies basic plasma processes, e.g. Landau damping. Connecting these two points of view would also contribute to addressing the fundamental question whether dissipation in plasmas is uniform or intermittent \citep{vaivads2016turbulence} and deserves a dedicated, future study. Moreover, kinetic plasma turbulence excites in a very complex way an entire ensemble of genuinely kinetic degrees of freedom, i.e. those related to velocity--space structures in the particle VDF \citep{servidio2017magnetospheric, pezzi2018velocityspace}, thus driving free energy towards finer and finer scales in velocity space where it is dissipated through interparticle collisions \citep{pezzi2019protonproton}. The relation of the dissipation surrogates considered here with the enstrophy phase-space cascade will be the subject of a future work.

Finally, there are many necessary extensions to the current study. The present simulations varied collisionality, but for a particular set of field and plasma initial conditions, so the parametric dependence of the conclusions attained herein should be the subject of future work. Both the reconnection and turbulence simulations were 2D, and therefore 3D effects are not captured. The reconnection simulations studied here employed a small system size in which the protons do not fully couple back to the large-scale systems, so it is important to revisit the proton dissipation measures in larger system sizes. The effect of proton--electron collisions on energy-based parameters in turbulence should be addressed in future work. The turbulence simulations considered here addressed decaying turbulence, and comparisons to driven turbulence would be interesting.

\section*{Acknowledgements}
This manuscript stemmed from discussions at Vlasovia 2019: The Sixth International Workshop on the Theory and Applications of the Vlasov Equation; we acknowledge many helpful conversations at the workshop. The authors thank an anonymous referee for her/his suggestions, which significantly improved the quality of the paper.
Simulations here described have been partially performed on the Marconi cluster at CINECA (Italy), within the projects IsC53$\_$RoC-SWT and IsC63$\_$RoC-SWTB, on the Newton cluster at the University of Calabria (Italy). This research partially used resources of the National Energy Research Scientific Computing Center (NERSC), under Contract no. DE-AC02-05CH11231. This work used the Extreme Science and Engineering Discovery Environment (XSEDE), which is supported by National Science Foundation grant number ACI-1548562. Resources supporting this work were also provided by the NASA High-End Computing (HEC) Program through the NASA Advanced Supercomputing (NAS) Division at Ames Research Center. HL acknowledges support from NSF EPSCoR RII-Track-1 Cooperative Agreement OIA-1655280 and an NSF/DOE Partnership in Basic Plasma Science and Engineering via NSF grant PHY-1707247. JJ was supported by a NSF Atmospheric and Geospace Science Postdoctoral Fellowship (Grant No. AGS-2019828). PAC gratefully acknowledges support from NSF Grant PHY-1804428, NASA Grants NNX16AG76G and 80NSSC19M0146, and DOE Grant DE-SC0020294. CLV was partially supported by EPN projects: PIM-19-01, PII-DFIS-2019-01 and PII-DFIS-2019-04. LSV was funded by the Swedish Contingency Agency, grant 2016-2102, and by SNSA grant 86/20. SS acknowledges funding from the European Union Horizon2020 research and innovation programme under grant agreement No. 776262 (AIDA, www.aida-space.eu). VR is supported by US DOE grant DE-SC0019315. JMT acknowledges support from NSF Grant AGS-1842638. WHM is partially supported by the MMS Theory and Modeling team under NASA grant 80NSSC19K0565. 

\section*{Data availability}
The data underlying this article will be shared on reasonable request to the corresponding author.






\bsp	
\label{lastpage}
\end{document}